\documentclass[showpacs,preprintnumbers,amsmath,amssymb]{revtex4}
\date{March 2010}
\usepackage{epsfig}
\usepackage{latexsym}
\usepackage{amssymb}
\usepackage{booktabs}
\usepackage{graphicx}
\usepackage{color}

\newcommand{\be}{\begin{equation}}
\newcommand{\ee}{\end{equation}}
\newcommand{\ba}{\begin{eqnarray}}
\newcommand{\ea}{\end{eqnarray}}
\newcommand{\bi}{\begin{itemize}}
\newcommand{\ei}{\end{itemize}}
\newcommand{\tr}{{\rm Tr\,}}

\newcommand{\<}{\langle}
\renewcommand{\>}{\rangle}
\newcommand{\eq}{Eq.~}

\newcommand{\la}{\label}

\newcommand{\Gw}{G_{\rm w}}
\newcommand{\hatp}{\hat{p}}
\newcommand{\circp}{p^{^{\!\!\!\circ}}}
\newcommand{\circq}{q^{^{\!\!\!\circ}}}
\newcommand{\hatk}{\hat{k}}

\renewcommand{\vec}[1]{\boldsymbol{#1}}

\begin{document}
\preprint{MITP/13-061 ~~~~ HIM-2013-06}
\title{Antiscreening of the Amp\`ere force in QED and QCD plasmas}

\author{Bastian Brandt}
\affiliation{Institut f\"ur theoretische Physik, Universit\"at Regensburg, D-93040 Regensburg}

\author{Anthony Francis, Harvey~B.~Meyer}
\affiliation{PRISMA Cluster of Excellence,
Institut f\"ur Kernphysik and Helmholtz~Institut~Mainz,
Johannes Gutenberg-Universit\"at Mainz,
D-55099 Mainz, Germany\vspace{0.2cm}}

\date{\today}

\begin{abstract}
The static forces between electric charges and currents are modified at the loop level by the presence of a plasma.
While electric charges are screened, currents are not. The effective coupling constant at long distances
is enhanced in both cases as compared to the vacuum, and by different amounts, a clear sign that
Lorentz symmetry is broken. We investigate these effects quantitatively, first in a QED plasma
and secondly using non-perturbative simulations of QCD with two light degenerate flavors of quarks.
\end{abstract}


\maketitle

\section{Introduction}

The properties of Abelian and non-Abelian plasmas are the subject of
intensive experimental and theoretical investigation.
See~\cite{Arnold:2012dqa} for a recent overview of the status of
research on the quark-gluon plasma, and~\cite{Thoma:2008gh} for a
comparative discussion of Abelian and non-Abelian plasmas.
Electromagnetic probes play an important role in heavy ion collision
experiments, see for instance the
recent~\cite{Gale:2012xq,Shen:2013vja}. In particular, it is important
to understand the way in which electromagnetic interactions are
modified by the presence of a thermal medium. While calculating
dynamical
properties~\cite{Brandt:2012jc,Ding:2010ga,Amato:2013naa,Ghiglieri:2013gia}
is crucial to predicting experimental observables in heavy-ion
collisions, here we will concentrate on static (time-independent)
aspects of the electromagnetic force in a plasma.

It is well-known that the electric force is screened by the quasi-free
charges of a plasma. This phenomenon of Debye screening occurs already
in a plasma described by classical physics. The static force between
currents however remains unscreened. The coupling constants of both
forces are in fact enhanced at long distances as compared to the
vacuum.  In the vacuum, the intensity
of the Coulomb force and the Amp\`ere force are related, a consequence
of Lorentz symmetry~\cite{BerkeleyEM}. In a medium such as a plasma,
they may have a different intensity. They receive a factor
$(1+e^2\kappa_\ell)$ and $(1+e^2\kappa_t)$ respectively, where both
$\kappa_\ell$ and $\kappa_t$ are found to be positive and
$\kappa_\ell>\kappa_t$. These coefficients are determined by the
long-wavelength behavior of the static polarization tensor. An
important point is that since the renormalized electromagnetic
coupling is defined conventionally at long distances in the vacuum
state (zero temperature), it is necessary to subtract the vacuum
polarization tensor in order to obtain the medium-induced modification
of the long-range force between renormalized charges and currents.

The enhancement of the effective coupling constant is perhaps
physically more relevant in the case of currents, since the Amp\`ere
force remains unscreened.  The enhancement of the latter can be
understood intuitively, by noting that free charges that happen to be
moving parallel to the external current line are attracted to it, thus
reinforcing it, while free charges moving antiparallel are
repelled. It is thus fair to speak of a medium-induced `antiscreening'
of the Amp\`ere force. This picture is corroborated by the fact that
the quantity $\kappa_t$, which quantifies the enhancement of the
force, also enters the constitutive equation of the electric current
at second order in a `hydrodynamic' description, $\vec j = e\kappa_t
\nabla\times \vec B \,+$ other terms~\cite{Hong:2010at}.

As a potential QED application, consider the early universe in its
first few minutes.  There is a period where electrons and positrons,
which are still relativistic, dominate the energy density along with
photons, with comparatively few nucleons present. The protons can then
be considered as slow-moving probes of the relativistic QED
plasma. They interact at long distances via a Debye-screened
potential. Its coefficient is modified by a factor
$(1+e^2\kappa_\ell)$. Note, however, that at the same order one also
has to consider the two-photon exchange amplitude.

A coefficient analogous to $\kappa_t$ has been introduced in the
constitutive equation of shear stress~\cite{Baier:2007ix}, where it is
denoted by $\kappa$. It multiplies a term that vanishes in flat space.
It has been computed in the SU($N$) gauge theory at weak coupling in
leading order~\cite{Romatschke:2009ng} and has recently been addressed
non-perturbatively using lattice Monte-Carlo
simulations~\cite{Philipsen2013}.

After deriving the physics interpretation of $\kappa_\ell$ and
$\kappa_t$, we recall the known one-loop formulae for the polarization
function in QED and extract the quantities $\kappa_\ell$ and
$\kappa_t$ (section \ref{sec:theo}). The QCD case is simply obtained
by multiplying the results by the number of colors carried by the
quarks.  In section \ref{sec:num} we then proceed to calculate for the
first time the polarization function in the deconfined phase of QCD.
We use Monte-Carlo simulations of lattice QCD with two flavors of
quarks (with the O($a$) improved Wilson action).

\section{Theory\la{sec:theo}}
In this section we define the correlation functions relevant to the 
calculation of the coefficients $\kappa_\ell$ and $\kappa_t$.
We give their interpretation in subsection \ref{sec:interpr}. 
In subsection \ref{sec:constitu}, we review the role of $\kappa_t$ as a 
second order coefficient in the constitutive equation of the electromagnetic 
current. Finally, we review the one-loop results for the 
polarization tensor and extract analytic representations for 
$\kappa_\ell$ and $\kappa_t$ useful at small and at large fermion masses.

\subsection{Definitions}

We work in the Euclidean field theory.
The vector current is defined as $j_\mu(x)=\bar\psi(x)\gamma_\mu\psi(x)$, where the Dirac matrices
are all hermitian and satisfy $\{\gamma_\mu,\gamma_\nu\}=2\delta_{\mu\nu}$. 
We define the polarization tensor as
\be\la{eq:PolTens}
\Pi_{\mu\nu}(q) \equiv \int d^4x \, e^{iq\cdot x} \<j_\mu(x) j_\nu(0)\>.
\ee
In the vacuum, it is purely transverse,
\ba
\Pi^{\rm vac}_{\mu\nu}(q) &=& P_{\mu\nu}(q) q^2 \Pi^{\rm vac}(q^2),
\\
P_{\mu\nu}(q) &=& q_\mu q_\nu/q^2 - \delta_{\mu\nu}.
\ea
With these conventions, the spectral function
\be\la{eq:rhoGG}
\rho^{\rm vac}(s) \equiv -2 \;{\rm Im}\, \Pi^{\rm vac}(-s -i\epsilon)\qquad  (s\geq 0)
\ee
is non-negative for a flavor-diagonal correlator. 
For the electromagnetic current, it is related to the $R$ ratio via
\be \la{eq:rhoR}
\rho^{\rm vac}(s) =\frac{R(s)}{6\pi},
\qquad
R(s) \equiv  \frac{\sigma(e^+e^-\to {\rm hadrons})}
 {4\pi \alpha(s)^2 / (3s) } .
\ee
The denominator is the treelevel cross-section $\sigma(e^+e^-\to\mu^+\mu^-)$
in the limit $s\gg m_\mu^2$, and we have neglected QED corrections.

At finite temperature $T\equiv 1/\beta$, the tensor decomposition reads
\be
\Pi_{\mu\nu}(q) = P^L_{\mu\nu}(q)\, \Pi_L(q_0^2,\vec q^2) + P^T_{\mu\nu}(q)\, \Pi_T(q_0^2,\vec q^2),
\ee
where $P^T_{00}(q)=0=P^T_{0i}(q)=P^T_{i0}(q)$ and $P^T_{ij}(q)= q_iq_j/\vec q^2 - \delta_{ij}$,
while $P^L_{\mu\nu}(q) = P_{\mu\nu}(q) - P^T_{\mu\nu}(q) $.

We define the `matter' part of the polarization tensor via
\ba
\Pi_{\mu\nu}(q) & \equiv & \Pi^{\rm vac}_{\mu\nu}(q) + \Pi_{\mu\nu}^{\rm mat}(q),
\\
\Pi_L^{\rm mat}(q)  & \equiv & \Pi_L(q) - q^2 \Pi^{\rm vac}(q^2), 
\\
\Pi_T^{\rm mat}(q)  &\equiv & \Pi_T(q) - q^2 \Pi^{\rm vac}(q^2).
\ea  
With these conventions, following~\cite{Hong:2010at} we define
\ba
\kappa_t &=& -\left. \frac{\partial}{\partial (q_3^2)} \Pi_{11}^{\rm mat}(q_3\vec e_3) \right|_{q_3=0}
 = \left. \frac{\partial}{\partial (\vec q^2)} \Pi^{\rm mat}_T(0,\vec q^2)\right|_{\vec q=0},
\\
\kappa_\ell &=& -\left. \frac{\partial}{\partial (q_3^2)} \Pi_{00}^{\rm mat}(q_3\vec e_3) \right|_{q_3=0}
 = \left. \frac{\partial}{\partial (\vec q^2)} \Pi^{\rm mat}_L(0,\vec q^2)\right|_{\vec q=0}.
\ea
We now turn to the interpretation of these quantities.

\subsection{Interpretation of $\kappa_\ell$ and $\kappa_t$ \la{sec:interpr}}

Consider an $SU(3)\times U(1)$ vector gauge theory with quarks as the
only matter fields.  Let $S$ be the action of quarks (described by the
field $\psi(x)$), gluons ($A_\mu(x)$) and photons ($B_\mu(x)$).  The
partition function of the system coupled to a classical
electromagnetic source $J_\mu(x)$ (it satisfies  $\partial_\mu J_\mu(x)=0$) reads
\be
Z[J] = \int D[\bar\psi] D[\psi] \,D[A]\,D[B]\; \exp\left(-S-ie_0 \int d^4x\; B_\mu(x) J_\mu(x)\right).
\ee
Expanding the free energy $F[J] = - \frac{1}{\beta}\log Z[J]$ in $J$, one obtains for the quadratic term
\be
F^{(2)}[J] = \frac{e_0^2}{2\beta} \int d^4x \int d^4 y\; \<B_\mu(x) B_\nu(y)\>_{J=0} \; J_\mu(x) J_\nu(y).
\ee 

\subsubsection{Free energy of electromagnetic sources at zero-temperature}
Consider first the situation at zero temperature.
In one loop approximation in the electromagnetic coupling $e$, the photon propagator is such that 
\be
\int d^4x \;J_\mu(x)\;\<B_\mu(x) B_\nu(y)\>_{J=0}  = 
\int d^4x \; J_\mu(x)\; \int \frac{d^4q}{(2\pi)^4} \; \frac{\delta_{\mu\nu}~e^{iq(x-y)}}{q^2(1-e_0^2\Pi^{\rm vac}(q^2))}.
\ee
For a static source of the form 
\be\la{eq:staticQ}
J_0(x) = Q_1 \delta^{(3)}(\vec x) + Q_2 \delta^{(3)}(\vec x - \vec r), 
\qquad \qquad J_i(x)=0,
\ee
$F^{(2)}[J] $ can be interpreted as the static potential between 
two static U(1) charges.
As a check that one recovers the familiar expression, for the source (\ref{eq:staticQ}) the
quadratic contribution to the free energy becomes 
\be
F^{(2)}(r) = V(r) = Q_1 Q_2 
\int \frac{d^3\vec q}{(2\pi)^3} \; \frac{e_0^2}{ 1-e_0^2\Pi^{\rm vac}(\vec q^2)} \frac{e^{i\vec q\cdot \vec r}}{\vec q^2} + \dots
\ee
where the remaining terms are independent of $\vec r$.
The renormalization of the electric charge leads to the substitution
\be
\frac{e_0^2}{1-e_0^2\Pi^{\rm vac}(\vec q^2)} = \frac{e^2}{1-e^2[\Pi^{\rm vac}(\vec q^2)-\Pi^{\rm vac}(0)]} 
\ee
at one-loop accuracy. We have thus recovered the familiar expression of the Coulomb potential,
modified by the vacuum polarization~\cite{Weinberg:1995mt}, the so-called Uehling potential.

\subsubsection{Generalization to finite temperature}

In momentum space, the $(00)$ component of the static (i.e.\ $q_0=0$) photon propagator is given at one loop by 
\be
\int d^4x\; e^{-i\vec q(\vec x-\vec y)}\; \<B_0(x) B_0(y)\>_{J=0} = \frac{1}{\vec q^2 - e_0^2\Pi_L(0,\vec q^2)}.
\ee
Since the electric charge is conventionally renormalized at zero temperature, the free energy added by the presence
of the two static leptons is given by 
\ba
F_l^{(2)}(r) &=& Q_1 Q_2
\int \frac{d^3\vec q}{(2\pi)^3} \;\frac{e_0^2}{\vec q^2 - e_0^2\Pi_L(0,\vec q^2)} \; e^{i\vec q\cdot \vec r}
\\ &=& Q_1 Q_2
\int \frac{d^3\vec q}{(2\pi)^3} \;
\frac{e^2}{1 - e^2[\Pi^{\rm mat}_L(0,\vec q^2)/\vec q^2 + \Pi^{\rm vac}(\vec q^2) - \Pi^{\rm vac}(0)]} \; 
\frac{e^{i\vec q\cdot \vec r}}{\vec q^2}.
\la{eq:F2gen}
\ea
Given the expansion at small momenta
\be
\Pi^{\rm mat}_L(0,\vec q^2) =-\chi_s + \kappa_l \vec q^2 + \dots, 
\ee
and $\Pi^{\rm vac}(\vec q^2) - \Pi^{\rm vac}(0)= {\rm O}(\vec q^2)$,
the Coulomb potential is screened at long distances, with a screening mass given by 
\be
m_{\rm el}^2 = e^2\chi_s .
\ee
Explicitly, the interaction free energy is given at long distances by
\be
F_l^{(2)}(r) =  \frac{e^2}{1-e^2\kappa_l}\; \frac{Q_1Q_2 \,e^{-m_{\rm el}r}}{4\pi r}.
\ee

Consider now the stationary source
\be
iJ_3(x) = I_1 \delta^{(2)}(\vec x_\perp) +I_2 \delta^{(2)}(\vec x_\perp+\vec R_\perp), \qquad \vec x_\perp\equiv(x_1,x_2),
\ee
where the other components of $J_\mu$ vanish. It describes two long wires along the $\vec e_3$ direction (which we take to be 
length $L_3$), separated by a distance $|\vec R_\perp|$. The currents $I_1$ and $I_2$ correspond to the charge, measured in units of $e$,
that flows per unit time along the $\vec e_3$ direction.
Then, up to $|\vec R_\perp|$-independent terms, 
\ba
F_t^{(2)}(R_\perp) &=& -I_1 I_2 L_3 \int \frac{d^2\vec q_\perp}{(2\pi)^2} \;
 \frac{e_0^2}{\vec q_\perp^2 -e_0^2 \Pi_T(\vec q_\perp^2)}\; e^{i\vec q_\perp \cdot\vec R_\perp}
\\ &=& -I_1 I_2 L_3 \int \frac{d^2\vec q_\perp}{(2\pi)^2} \;
 \frac{e^2}{ 1 -e^2 [\Pi^{\rm mat}_T(\vec q_\perp^2)/\vec q_\perp^2 + \Pi^{\rm vac}(\vec q_\perp^2) - \Pi^{\rm vac}(0)] }\;
  \frac{e^{i\vec q_\perp \cdot\vec R_\perp}}{\vec q_\perp^2}
\ea
Thus we see that the transverse channel $\Pi_T$ modifies Amp\`ere's $1/R_\perp$ force law.
At long distances, the modification is given by 
\be
-\frac{1}{L_3} \frac{\partial F_t^{(2)}}{\partial R_\perp}
= - \frac{e^2}{1 -e^2 \kappa_t} \cdot \frac{I_1I_2}{2\pi R_\perp}.
\ee
Thus a positive value of $\kappa_t$ corresponds to an enhancement of
the strength of the $1/R_\perp$ force law, as compared to the Amp\`ere force in the vacuum.

\subsection{Constitutive equation of the electromagnetic current\la{sec:constitu}}

The equation of continuity for the electric charge, $\dot\rho + \nabla\cdot \vec j =0$,
can be used to describe the time evolution of a non-equilibrium initial charge distribution.
This `hydrodynamic' description applies on long distance and time scales.
It needs to be supplemented by a constitutive equation in order to be predictive.
At lowest order in a derivative expansion, $e\vec j = \sigma \vec E -eD\nabla \rho$, 
with the conductivity $\sigma$ and the diffusion coefficient related by $\sigma = \chi_s D$.
At the next order, the most general allowed terms are of the form~\cite{Hong:2010at}
\be\la{eq:consti}
(1 + \tau_J \partial_t) e\vec j =  - e D\nabla \rho + \sigma  \vec E+ \kappa_B \nabla\times \vec B.
\ee
Note that in our conventions, for the QCD plasma
 $\sigma$ and $\kappa_B$ are both of order $e^2$ in the electromagnetic coupling.
Through the linear response formalism, one shows~\cite{Hong:2010at}
that the coefficient multiplying the the curl of $\vec B$ is directly
related to the quantity $\kappa_t= \kappa_B / e^2$ introduced above from the medium-induced part
of the polarization function $\Pi_T^{\rm mat}$.

Equation (\ref{eq:consti}) confirms the interpretation that the
coefficient $\kappa_t>0$ corresponds to an antiscreening of electric
currents in the medium. Indeed, suppose an external current
$I_0=I_{\rm ext}$ flowing along a straight line is applied to the
system, creating circular magnetic field lines around it. If we choose
a transverse disc $S$ (with an edge $\partial S$) centered on it, the
response electric current $e\int_S d\sigma \cdot \vec j$ flowing
through $S$ is given by $I_1=\kappa_B \int_{\partial S} d\vec \ell
\cdot \vec B$. This current enhances the circulation of the magnetic
field along $\partial S$ by an amount equal to $I_1$, which in turn
induces a current $I_2$.  This amounts to a geometric series
$\sum_{k=0}^\infty I_k$, so that the net effect is to enhance $I_0$ by
a factor $1/(1-\kappa_B)$.

\subsection{Leading-order perturbative results\la{sec:LOp}}

We now give one-loop results for the polarization tensor and its small
$\vec q^2$ expansion in a pure QED plasma with one Dirac fermion of
mass $m$ carrying a charge $e$.  The expressions for QCD are obtained
by summing the contribution of each quark flavor and multiplying the
result by the number of colors $N_c$.

In QED at leading order, the vacuum polarization reads (see for instance~\cite{Weinberg:1995mt})
\ba
\Pi^{\rm vac}(q^2)-\Pi^{\rm vac}(0) &=& \frac{1}{2\pi^2}\int_0^1 dx \, x(1-x)\log\left[1+\frac{q^2x(1-x)}{m^2}\right].
\ea
Two independent components of the thermal contribution read~\cite{Kapusta:2006pm},
with $E_{\vec k} = \sqrt{\vec k^2+m^2}$ and $n_F(E) = 1/(e^{\beta E}+1)$, 
\ba\la{eq:Pi00_1loop}
\Pi_{00}^{\rm mat}(q) &=& \frac{2}{\pi^2} {\rm Re}\int_0^\infty {k^2dk}\, \frac{n_F(E_{\vec k})}{E_{\vec k}}
\left[ 1 + \frac{4E_{\vec k}^2 - 4iE_{\vec k}q_0 - q_0^2 - \vec q^2}{4k|\vec q|}\, \log\frac{R_+}{R_-}\right],
\\
\la{eq:Pimumu_1loop}
\Pi^{\rm mat}_{\mu\mu}(q) &=& \frac{4}{\pi^2}{\rm Re}\int_0^\infty k^2 dk\, \frac{n_F(E_{\vec k})}{E_{\vec k}}
 \left[1+\frac{2m^2-q_0^2-\vec q^2}{4|\vec q|k} \log\frac{R_+}{R_-}\right],
\nonumber
\\
R_{\pm} &=& {q_0^2+\vec q^2 \pm 2|\vec q|k +2iq_0 E_{\vec k}}.
\ea
From here one obtains the longitudinal and transverse part of the polarization tensor via
\be
\Pi_L(q_0^2,\vec q^2) = - \Pi_{00}(q), \qquad \qquad 
\Pi_T(q_0^2,\vec q^2) = \frac{1}{2}\left( \Pi_{00}(q) - \Pi_{\mu\mu}(q)\right).
\ee

\begin{figure}[t]
\centerline{\includegraphics[width=0.5\textwidth,angle=0]{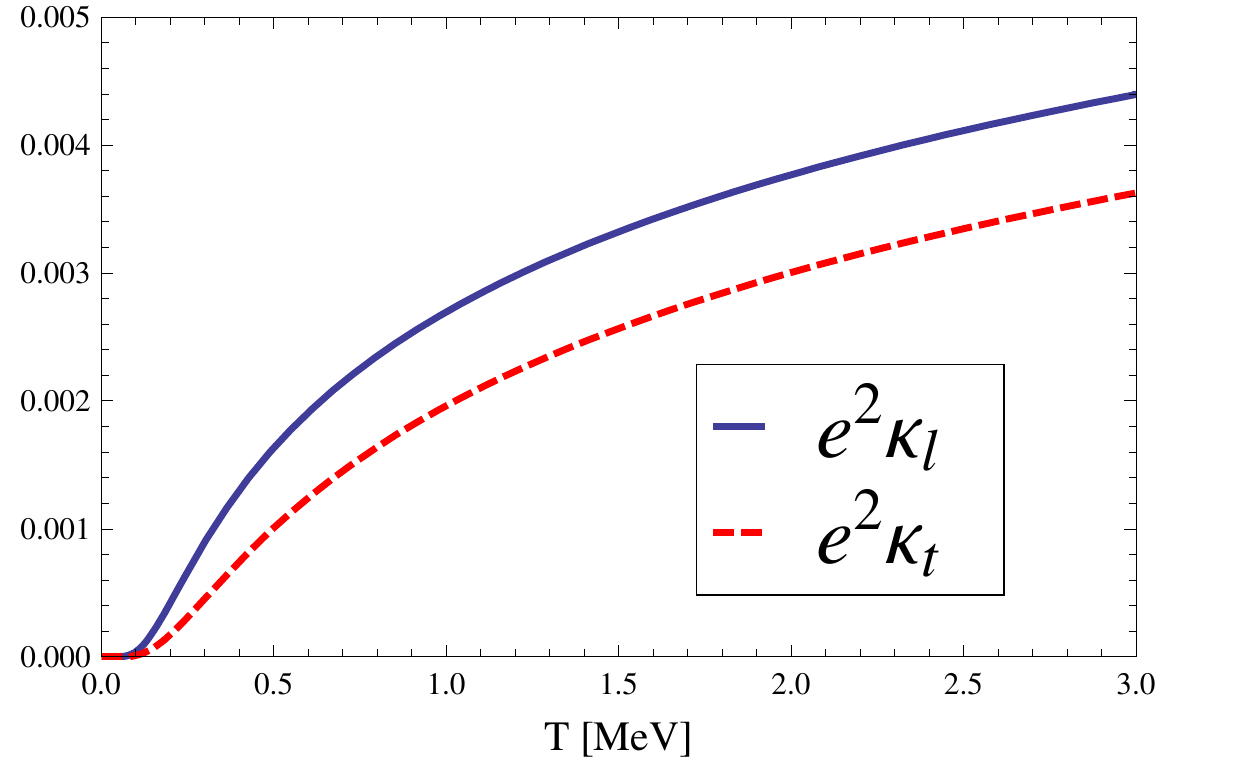}}
\caption{The coefficients $e^2\kappa_\ell$ and $e^2\kappa_t$ at one loop in QED, with 
the fermion mass set to $m_e=0.511{\rm MeV}$ and $e^2=4\pi/137$.}
\label{fig:kaplkaptQED}
\end{figure}

The expression for the fermion number susceptibility reads
\be
\Pi_{00}^{\rm mat}(q=0)= -\Pi_L(0,0) = -\Pi^{\rm mat}_L(0,0)  = \chi_s = \frac{2}{\pi^2} \int_0^\infty \frac{dk}{E_{\vec k}}\, 
n_F(E_{\vec k})\; (k^2+E_{\vec k}^2)
= -\frac{2m^2}{\pi^2} \sum_{n\geq 1} (-1)^n \; K_2(n\, \beta m).
\ee
In the massless limit, we have $\chi_s = T^2/3$.

In the transverse channel, we obtain from (\ref{eq:Pi00_1loop}--\ref{eq:Pimumu_1loop})
\be
\kappa_t = \frac{1}{3\pi^2}\int_0^\infty \frac{dk}{E_{\vec k}}\; n_F(E_{\vec k})
 = -\frac{1}{3\pi^2} \sum_{n=1}^\infty (-1)^n \,K_0(n\,\beta m).
\ee
For small $\beta m$, we find the expansion (see appendix \ref{sec:apdxPT})
\be\la{eq:kappabmassless}
\kappa_t = \frac{1}{6\pi^2} \left(\log\left({1}/{b}\right)  - \gamma_{\rm E}\right) + {\rm O}(b^2\log(1/b)),
\qquad b \equiv \frac{m}{\pi T}.
\ee
There is an infrared divergence coming from the $T=0$ contribution. In QCD, this divergence 
is driven by the light pions rather than by the light quarks, but the infrared sensitivity 
is also present.

It is convenient to consider the difference $\kappa_\ell - \kappa_t $, since 
it is infrared safe, unlike  $\kappa_\ell$ and $\kappa_t$ individually. We find 
\be\la{eq:kp-kbexp}
\kappa_\ell - \kappa_t = -\frac{1}{6\pi^2} \int_0^\infty dp \;  n_F'(E_{\vec p})
   = -\frac{\beta m}{6\pi^2} \sum_{n=1}^\infty (-1)^n\, n \, K_1(n\,\beta m),
\ee
where  $n_F'(E) \equiv \frac{dn_F}{dE}$.
At small $\beta m$, its expansion is
\be
\kappa_\ell - \kappa_t = \frac{1}{12\pi^2} \left(1 - {\frac{7}{4}}\zeta_3\, b^2 + {\frac{93}{32}}\zeta_5\, b^4
 + {\rm O}(b^6)\right),
\ee
and it would not be difficult to extend the series to higher orders.
Note that  $\kappa_\ell - \kappa_t$ amounts to a small number compared 
to the fermionic contribution to the entropy, $s/T^3 = 7\pi^2/45$ in the massless case.

Finally, as a numerical application, we display the value of the
coefficients $\kappa_\ell$ and $\kappa_t$ in QED as a function of the
temperature in Fig.\ \ref{fig:kaplkaptQED}. If we think about the
early universe, we note that around the time when free
neutrons start to disappear by $\beta$ decay ($T\simeq0.8{\rm MeV}$,
see for instance the recent review~\cite{Olive:2012ena}),
$\kappa_\ell$ and $\kappa_t$ represent a 2--3 permille effect on the
inter-proton electromagnetic force. By the time the nucleosynthesis chain
starts (around $T=0.1{\rm MeV}$), the effect really is tiny.

If we now consider the quark-gluon plasma, we expect the difference
$\kappa_\ell - \kappa_t$ to be well described by the leading
perturbative formulae above (with the suitable $N_c$ and $N_f$
factors) at sufficiently high temperature, since it is not
infrared-sensitive at leading order in perturbation theory. The value
of the individual coefficients $\kappa_\ell$ and $\kappa_t$ however
emerge from an interesting interplay between ultraviolet and infrared
physics. Indeed, without the vacuum subtraction, they would be
ultraviolet divergent, but on the other hand, the vacuum contribution
is infrared divergent for massless fermions. In QCD, this infrared
contribution is non-perturbative (it is dominated by pions), and
therefore we turn to lattice simulations in order to evaluate
$\kappa_\ell$ and $\kappa_t$ individually.

\section{Numerics\la{sec:num}}

\begin{figure}[t]
\centering
\includegraphics[width=.5\textwidth]{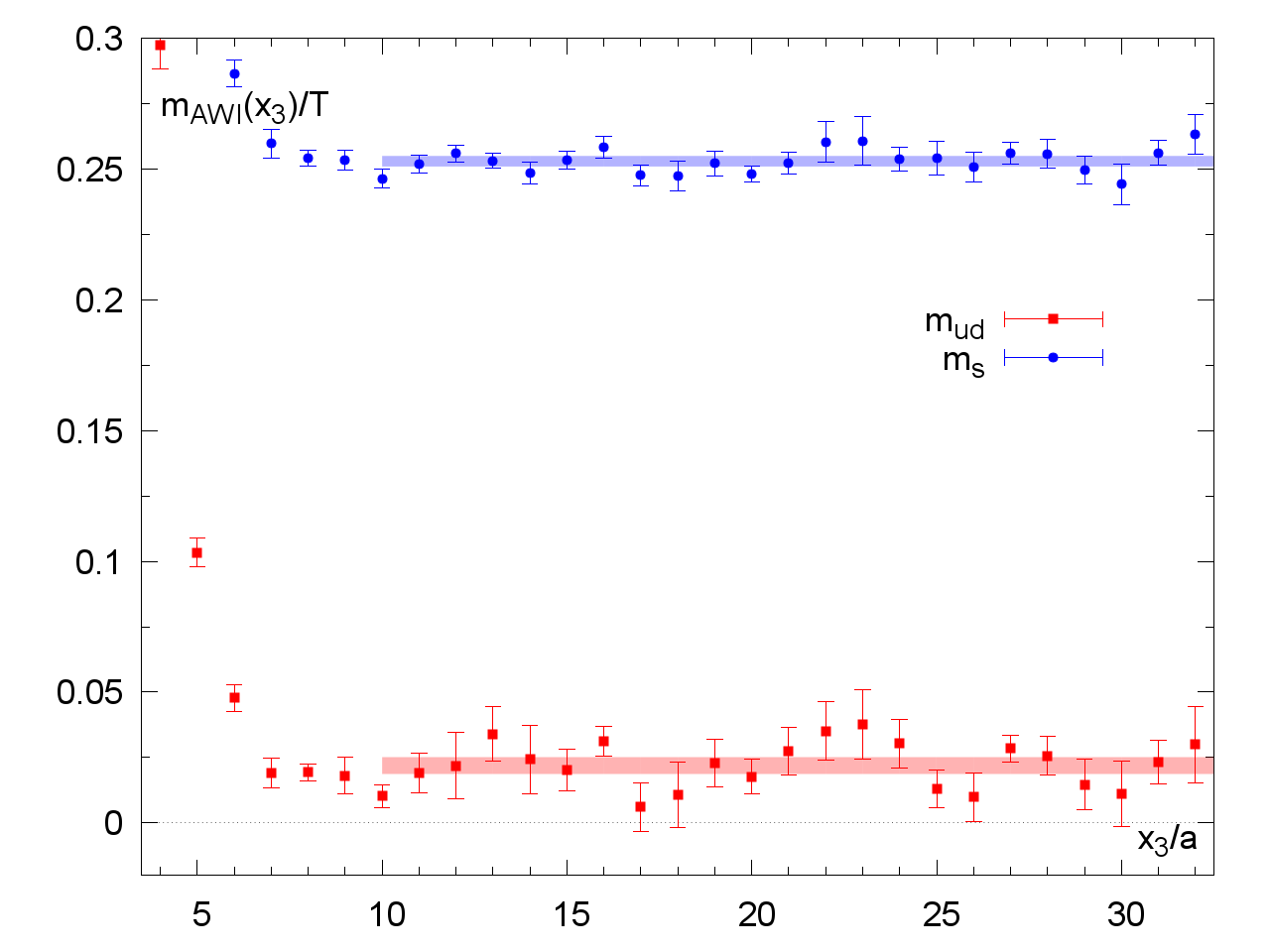}
\caption{{The PCAC quark masses in the $x_3$ direction as calculated 
on our finite temperature lattice ensemble of size $16\times64^3$. 
See the text for details on obtaining the quoted values $m/T$.
}}
\label{fig:quarkmass}
\end{figure}

In this section, we describe a numerical lattice QCD calculation of
the antiscreening coefficients $\kappa_l$ and $\kappa_t$, as well as
of the free energy of two static leptons for a general separation $r$.

All our numerical results were obtained on dynamical gauge
configurations with two mass-degenerate quark flavors.  The gauge
action is the standard Wilson plaquette action \cite{Wilson:1974sk},
while the fermions were implemented via the O($a$) improved Wilson
discretization with non-perturbatively determined clover coefficient
$c_{\rm sw}$ \cite{Jansen:1998mx}.  The configurations
were generated using the MP-HMC algorithm~\cite{Hasenbusch:2001ne,Hasenbusch:2002ai} in the
implementation of Marinkovic and Schaefer~\cite{Marinkovic:2010eg}
based on L\"uscher's DD-HMC package~\cite{CLScode}.  

We calculated correlation functions using the same discretization and
masses as in the sea sector on two lattice ensembles. One at zero
temperature and a lattice of size $128\times 64^3$ (labeled O7
in~\cite{Fritzsch:2012wq}) with a lattice spacing of
$a=0.0486(4)(5)$fm~\cite{Fritzsch:2012wq} and a pion mass of
$m_\pi=270$MeV, so that $m_\pi L = 4.2$.  The second at finite
temperature with a lattice of size $16\times 64^3$ with all bare
parameters identical to the zero-temperature ensemble.  The
zero-temperature ensemble was made available to us through the CLS
effort~\cite{CLS}, while the second ensemble was generated by us and
already presented in~\cite{Brandt:2012jc}. Note that choosing
$N_\tau=16$ yields a temperature of $T\simeq250$MeV. Based on
preliminary results on the pseudo-critical temperature $T_c$ of the
crossover from the hadronic to the high-temperature
phase~\cite{Brandt:2012sk}, the temperature can also be expressed as
$T/T_c\approx 1.2$.

In addition we calculated correlation functions on the same ensembles
with the bare quark mass tuned to match the physical strange quark
mass~\cite{Capitani:2011fg}. More precisely, the bare quark mass is
fixed at zero temperature by tuning the kaon mass to the value 
realized in nature.  According to perturbation theory, the most
relevant dimensionless parameter is $m/T$.  In order to facilitate a
comparison with the one-loop results, we therefore measure this
quantity directly using the partially conserved axial current (PCAC)
relation~\cite{Bochicchio:1985xa,Luscher:1996ug} in the
finite-temperature ensemble.  Since it is an operator identity, we are
free to evaluate the correlation functions of the axial current and
the pseudoscalar density in the longer spatial direction to evaluate
the quark mass. Specifically, we define
\be
m_{\rm AWI}(x_3)=\frac{1}{2}\sum_{(x_0,x_1,x_2)} \frac{\langle \partial^{\rm imp}_3
J^I_{3,5}(x)   J_5(0) \rangle  }{ \langle J_5(x_i) J_5(0)
\rangle}\quad,
\ee
where $J_5(x_i)$ and $J^I_{\mu,5}(x_i)=J_{\mu,5}(x_i) + ac_A \partial^{\rm imp}_\mu
J_5(x_i)$ denote the (isovector) pseudoscalar density and improved axial-vector
current, respectively. As `improved' lattice derivatives $\partial^{\rm imp}_\mu$ we use the 
higher-order difference scheme given in~\cite{Guagnelli:2000jw}, Eq.\ (2.18--2.19).
Here, for the improvement coefficient $c_A$ we
use the non-perturbatively determined value of \cite{DellaMorte:2005se}.
The extraction of the PCAC mass is illustrated in Fig.\ \ref{fig:quarkmass}. The quoted
values for $m/T$ were determined by fitting a constant to $m_{\rm AWI}(x_3)$ 
over the interval indicated by the band in Fig.\ \ref{fig:quarkmass}. 
The result was found to be stable under variations of the fit interval.
To renormalize the PCAC quark mass we use the
non-perturbatively determined renormalization factors $Z_A$ and $Z_P$
of the axial current and the pseudoscalar density of
\cite{Fritzsch:2012wq,DellaMorte:2008xb}, as well as the conversion factor from the 
Schr\"odinger functional scheme to the $\overline{\rm MS}$ scheme from~\cite{DellaMorte:2005kg,Fritzsch:2012wq}. 
Altogether the factor by which we multiply the bare PCAC mass 
to obtain the $\overline{\rm MS}$ mass at a renormalization scale of $\mu=2\,$GeV is $1.481(34)$.
The result as well as all relevant parameters are collected in Tab.\ \ref{tab:pars}.
From now on we simply refer to these $\overline{\rm MS}$ quark masses by $m$.  

\begin{table}[t]
\centering
\begin{tabular}{l@{~~~~~~}l}
\hline\hline\noalign{\smallskip}
$6/g_0^2$ & 5.50\\
$\kappa$ & 0.13671\\
$c_{SW}$ & 1.751496\\
\noalign{\smallskip}\hline\noalign{\smallskip}
$m_\pi$[MeV]   &  270\\
$Z_V$ & 0.768(5)\\
$a[\textrm{fm}]$ & 0.0486(4)(5) \\
$T_{N_\tau=16}$[MeV] & 253(4)  \\
\noalign{\smallskip}\hline\noalign{\smallskip}
$\overline m_{\rm ud}^{\overline{\rm MS}}(2\,{\rm GeV})/T$ &  0.0325(48)(7) \\ 
$\overline m_{\rm s}^{\overline{\rm MS}}(2\,{\rm GeV})/T$ &  0.3747(30)(86)     \\  
\noalign{\smallskip}\hline\hline
\end{tabular}
\caption{{The top block shows the bare lattice parameters, for more details on the $N_\tau=128$ and $N_\tau=16$ 
ensembles  see \cite{Fritzsch:2012wq,Brandt:2012jc,Brandt:2012sk}. 
The middle block summarizes the pion mass, the vector renormalization constant~\cite{DellaMorte:2005rd}, the lattice spacing~\cite{Fritzsch:2012wq}
and the corresponding temperature of our $N_\tau=16$ lattice calculation. 
In the bottom block we give the renormalized quark masses in units of temperature, where the first error
is from the bare axial Ward identity mass and the second from the renormalization factor.}}
\label{tab:pars}      
\end{table}

We implemented the vector correlation function
as a mixed correlator between the local and the conserved current.
In the following we will require the three correlation functions: 
\ba \la{eq:gcorr}
&G^{\rm bare}(x_0,g_0,T)\,\delta_{kl} &= - a^3 \sum_{(x_1,x_2,x_3)} \< J^c_k(x) J^l_\ell(0) \>, \quad k,l=1,2,3,\\  \la{eq:gtcorr}
&G_t^{\rm bare}(x_3,g_0,T)\,\delta_{ij} &= - a^3 \sum_{(x_0,x_1,x_2)} \< J^c_i(x) J^\ell_j(0) \>,\quad i,j=1,2,\\ \la{eq:glcorr}
&G_l^{\rm bare}(x_3,g_0,T) &= - a^3 \sum_{(x_0,x_1,x_2)} \< J^c_0(x) J^\ell_0(0) \>,
\ea
where
\ba
J_\mu^l(x) &=& \frac{1}{\sqrt{2}}\bar q(x) \gamma_\mu {\tau^3} q(x),
\\
\la{eq:Jdef}
J_\mu^c ( x ) &=& \frac{1}{2\sqrt{2}} \Big(\bar q ( x + a\hat\mu ) ( 1 + \gamma_\mu ) U_\mu^\dagger ( x ) {\tau^3} q ( x ) 
 - \bar q ( x )( 1 - \gamma_\mu ) U_\mu ( x ) {\tau^3} q ( x + a\hat\mu ) \Big).
\ea
Here $q$ represents a doublet of mass-degenerate quark fields and $\tau^3$ the diagonal Pauli
matrix acting on the flavor indices. The doublet can be interpreted as the (u,\,d) quarks
(which are treated fully dynamically) for the light mass case, while it can be interpreted as a
`partially quenched' (s,\,s$'$) doublet for the heavier case (i.e.\ their back-reaction on the 
thermal system is neglected).
We note that with this normalization of the current, 
the perturbative prediction for its two-point function 
is given by $N_c=3$ times the expressions given in section \ref{sec:LOp}.

We have renormalized the vector correlator using
\ba
G_{..}(x_i,T)= Z_V(g_0)\;  G_{..}^{\rm bare}(x_i,g_0,T) 
\ea
with the non-perturbative value of
$Z_V=0.768(5)$~\cite{DellaMorte:2005rd}.  We have not included O($a$)
contributions from the improvement term proportional to the derivative
of the antisymmetric tensor
operator~\cite{Luscher:1996sc,Sint:1997jx}. A quark-mass dependent
improvement term of the form $(1+b_V(g_0)am_q)$~\cite{Sint:1997jx} was
also neglected. These contributions should eventually be included to
ensure a smooth scaling behavior as the continuum limit is taken.
Here, our primary goal is to carry out the analysis on a single
lattice spacing.

\subsection{Correlator data and screening masses}

\begin{figure}[t]
\centerline{\includegraphics[width=.5\textwidth]{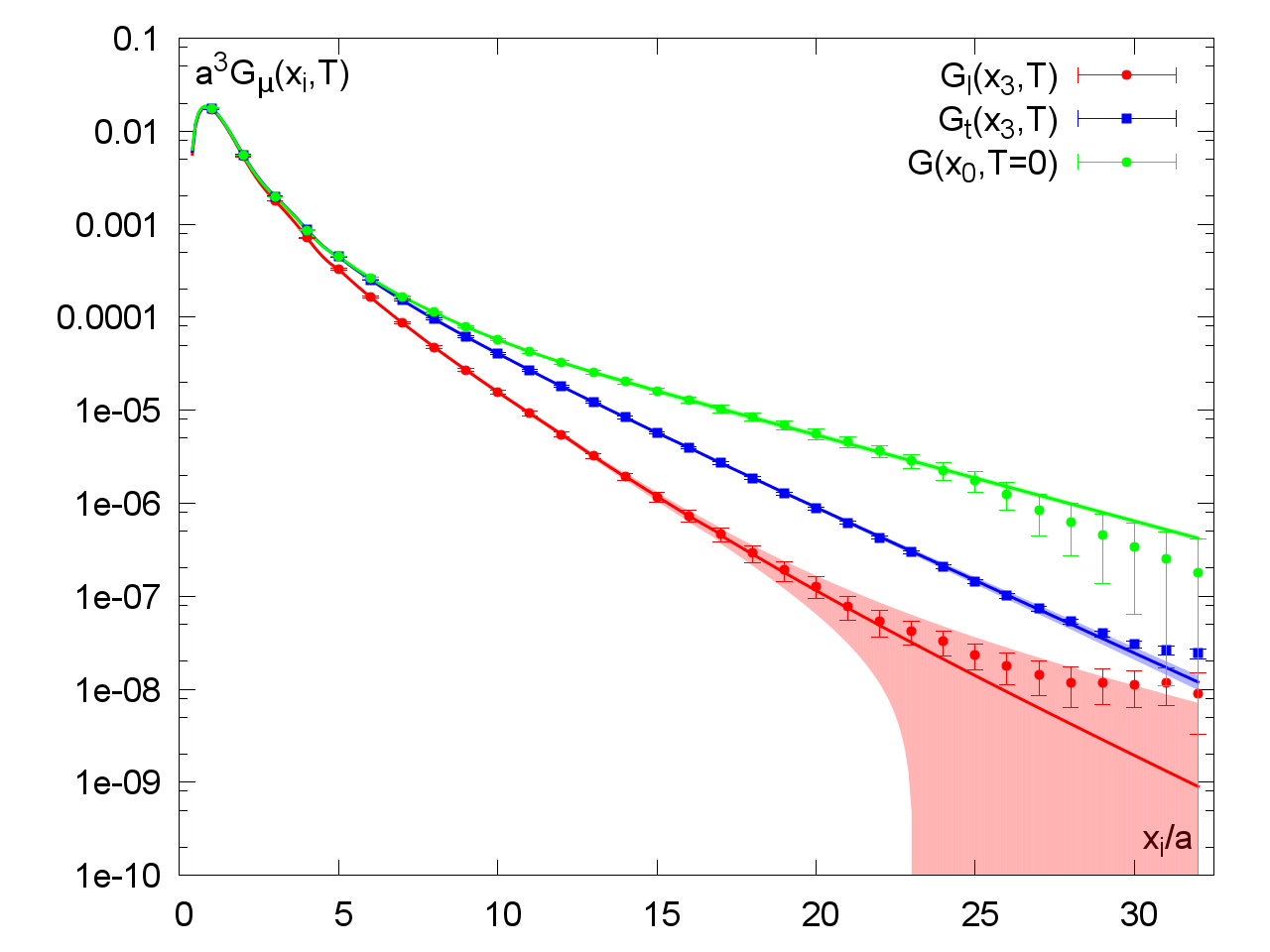}
\includegraphics[width=.5\textwidth]{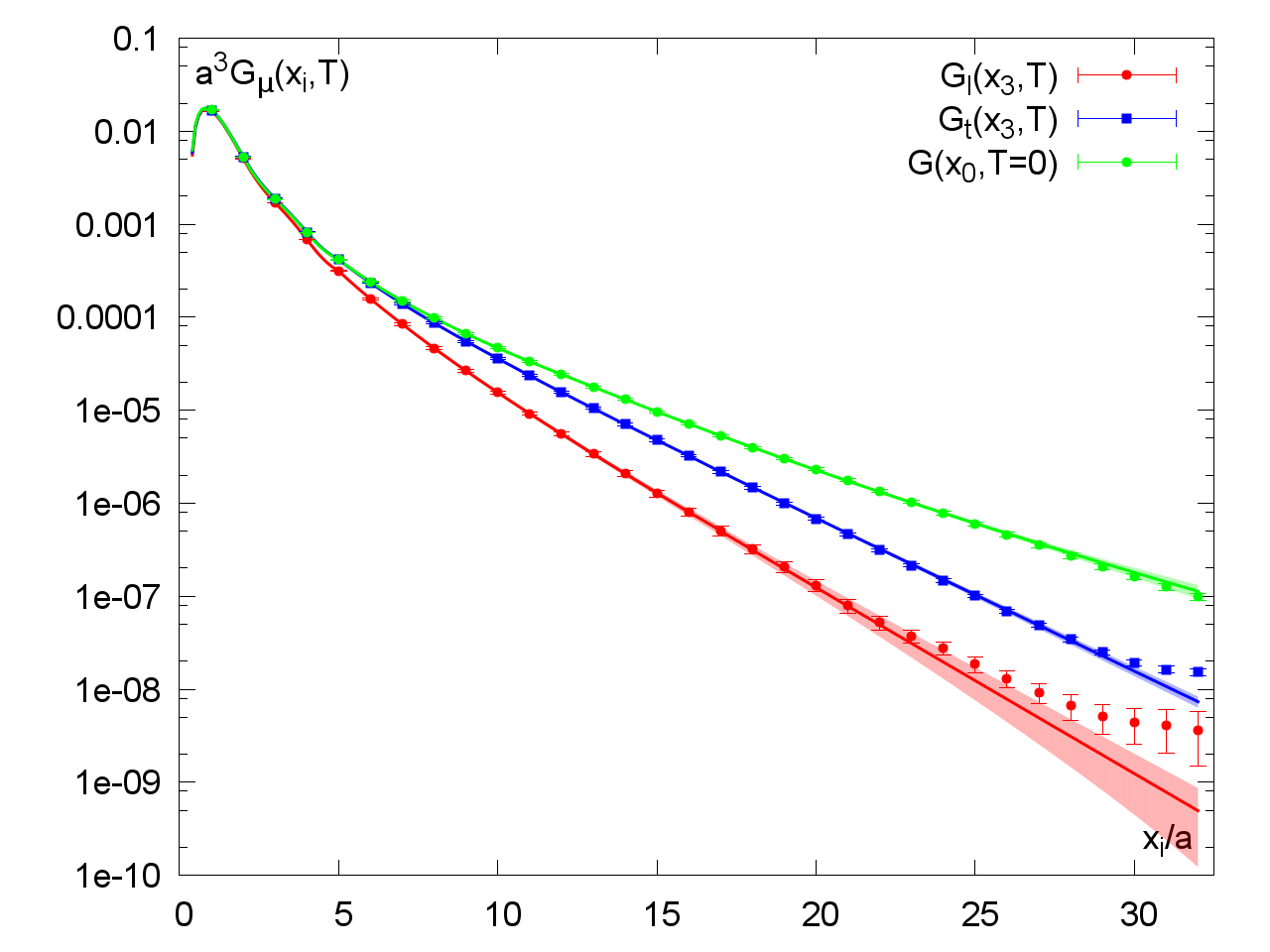}}
\caption{{Local-conserved vector correlation functions for the light quarks (left panel) 
and the strange quark (right panel). The shaded bands represent the correlators 
entering the computation of the antiscreening coefficients and the free energy.}}
\label{fig:correlators}
\end{figure}

Reaching our goal of computing the antiscreening coefficients and the
free energy requires the computation of three vector correlation
functions: the vector correlator at zero temperature
(Eq.\ \ref{eq:gcorr}) in the $x_0$ direction, as well as the individual
transverse (Eq.\ \ref{eq:gtcorr}) and longitudinal
(Eq.\ \ref{eq:glcorr}) vector correlators in the $x_3$-direction.

To treat the latter required integrals over the spatial vector
correlation functions without introducing unnecessarily large finite
lattice effects, we employ the position space representation introduced
in~\cite{Francis:2013fzp}. Here, the local-conserved correlator was
extrapolated with an exponential that decays with the lowest-lying
`mass'~\footnote{The energy level extracted in this way does not
  necessarily correspond to a stable vector particle.}.  This mass can
be fixed by fitting to the lattice data an Ansatz of the form
\be
G_{\rm Ansatz}(x_i)=\sum_{n=1}^2 |A_n|^2 e^{-m_n x_i},
\ee
for $x_i$ sufficiently below the half-lattice extent $N_{i}/2$ that
the `backward' propagating states give a negligible
contribution. 
In the zero temperature ensemble this
mass can be determined reliably by extracting it from a separate correlation
function, computed on the same configurations using smeared operators
at the source and sink~\cite{Francis:2013fzp,Capitani:2011fg}. 
This operator has greater overlap with the ground state and yields
more precise data. The mass parameter determined in this way is then
carried over to the local-conserved correlator and the corresponding
prefactor of the exponential $|A_1|^2$ is fitted to the lattice data 
 around $x_0=\beta/4$.  For the finite temperature
ensembles, we did not compute vector correlation functions using smeared-smeared operators
but fitted using the local-conserved data directly. However, due to
the better signal-to-noise ratio in the finite temperature case,
accurate results can nevertheless be obtained in this fashion.

\begin{figure}[t]
\centerline{\includegraphics[width=.5\textwidth]{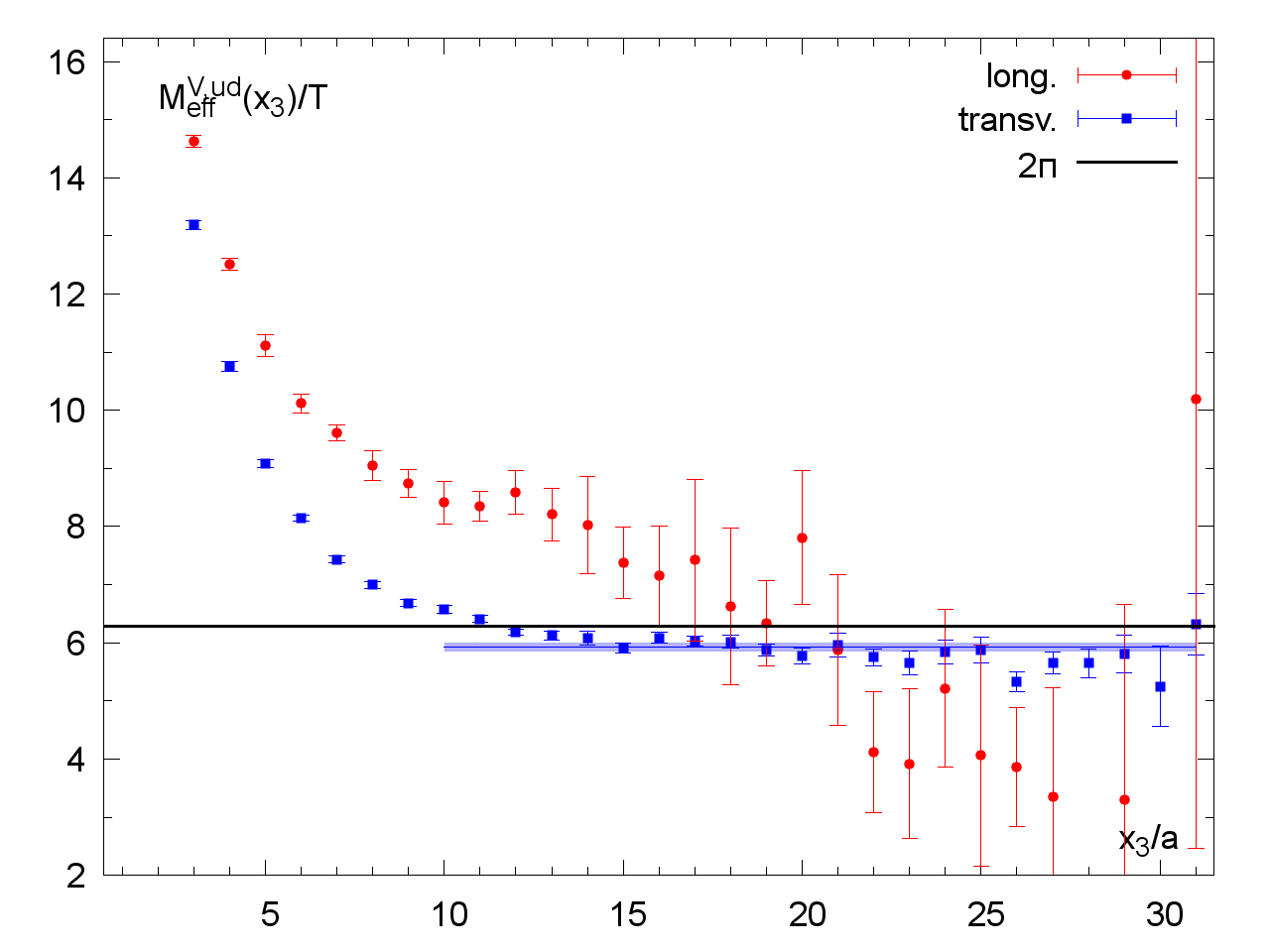}
\includegraphics[width=.5\textwidth]{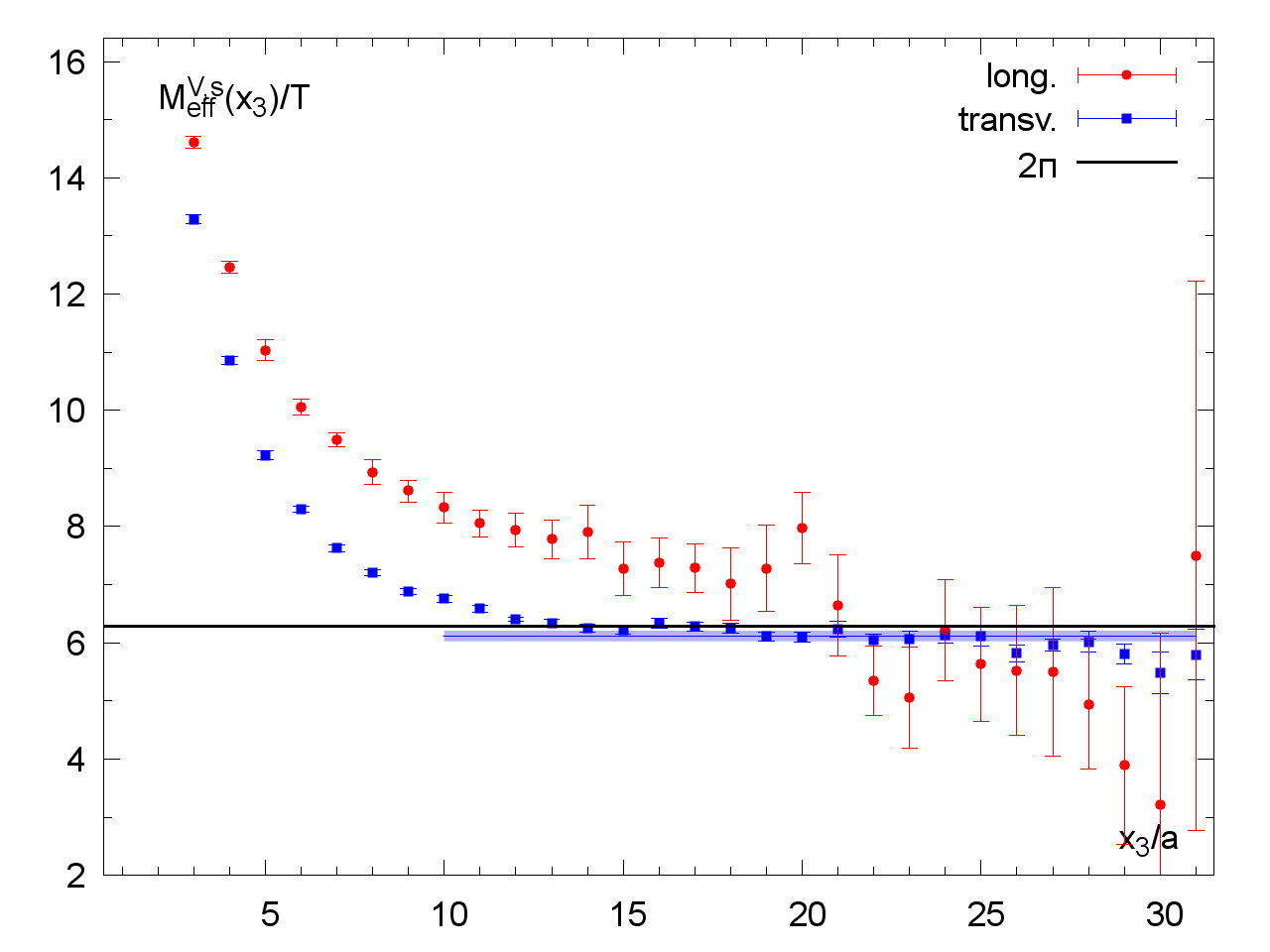}}
\caption{{The effective screening masses (defined by \eq (\ref{eq:rat})) for the longitudinal 
and the transverse channels. The left panel corresponds to the light quark case and the right 
panel to the strange quark. The band shows the result of the fit for the mass with its statistical error. }}
\label{fig:effmass}
\end{figure}

The raw lattice data as well as the resulting correlation functions
are shown as the colored shaded bands in Fig.\ \ref{fig:correlators},
where the error estimates were obtained via a jackknife procedure. 
For the light-quark longitudinal correlator, the relative errors
grow rapidly for $x_3/a\gtrsim 20$, but this 
has little impact on our determination of the antiscreening coefficients 
and the free energy of two static leptons.

As a byproduct of our study, we can investigate the lowest-lying
screening masses coupling to the isovector vector current. For that
purpose, in Fig.\ \ref{fig:effmass} we display the effective screening
masses, defined by the implicit equation
\be
\frac{G(x_3-a/2)}{G(x_3+a/2)} =
\frac{\cosh\big[M_{\rm eff}^V(x_3)({L}/{2}-(x_3-a/2))\big]}
     {\cosh\big[M_{\rm eff}^V(x_3)({L}/{2}-(x_3+a/2))\big]}.
\la{eq:rat}
\ee
This form corrects for the leading effects of the 
finite length of the $x_3$ direction.
In the transverse channel, we observe a convincing plateau, showing that we 
observe the asymptotic screening mass.
In the longitudinal channel, the correlator initially falls off with a substantially
higher exponent, but at distances beyond $x_3\approx \beta$, the effective exponent 
appears to decrease to a value below $2\pi T$. A possible interpretation is that the 
charge density operator couples only weakly to the lightest state in that channel, 
and therefore initially decays with a higher exponent. 
Given this behavior, we leave the extraction of a screening mass in the longitudinal channel 
for a future study with increased statistics and possibly improved spectroscopic methods.
In the transverse channel, we extract the following values for the screening mass,
\be
\frac{M^V_t}{2\pi T} = \left\{ \begin{array}{l@{~~~~}l}
0.943(17)   &  \textrm{light quarks}  
\\
0.973(21)   &  \textrm{strange quarks}.
\end{array}\right.
\ee
The quoted error contains an estimate of the uncertainty associated 
with choosing a fit interval.
We thus have rather convincing evidence that the screening mass lies
below $2\pi T$ for light quarks.  These results can be compared to the
perturbative prediction $M_t^V/(2\pi T) = 1 + 0.02980 g^2$ for two
massless flavors of quarks~\cite{Laine:2003bd}. Presumably the value
of the screening mass reaches values above $2\pi T$ at sufficiently
high temperatures. Interestingly, the charge density operator
(corresponding to the longitudinal channel) also plays a somewhat
special role in the perturbative analysis~\cite{Laine:2003bd}. It
would be worth revisiting this special case. In addition to the masses,
the coupling $|A_1|^2$ of the lightest screening state to the vector current is  of significant interest.
In the regime where the hierarchy $gT\ll \pi T$ applies, 
it is proportional to the wavefunction at the origin describing the bound state
of two fermions effectively of mass $\pi T$. We find, in the transverse channel,
\be
\frac{1}{T^3}|A_1|^2 = \left\{ \begin{array}{l@{~~~~}l}
6.05 \pm 0.78  &  \textrm{light quarks} 
\\
5.80 \pm 1.10  &  \textrm{strange quarks}.
\end{array}\right.
\ee
A future comparison with perturbative calculations would be complementary to the 
comparison of the screening masses.

\subsection{The antiscreening coefficients $\kappa_t$, $\kappa_l$ and $(\kappa_l-\kappa_t)$}

\begin{figure}[t]
\centerline{\includegraphics[width=.5\textwidth]{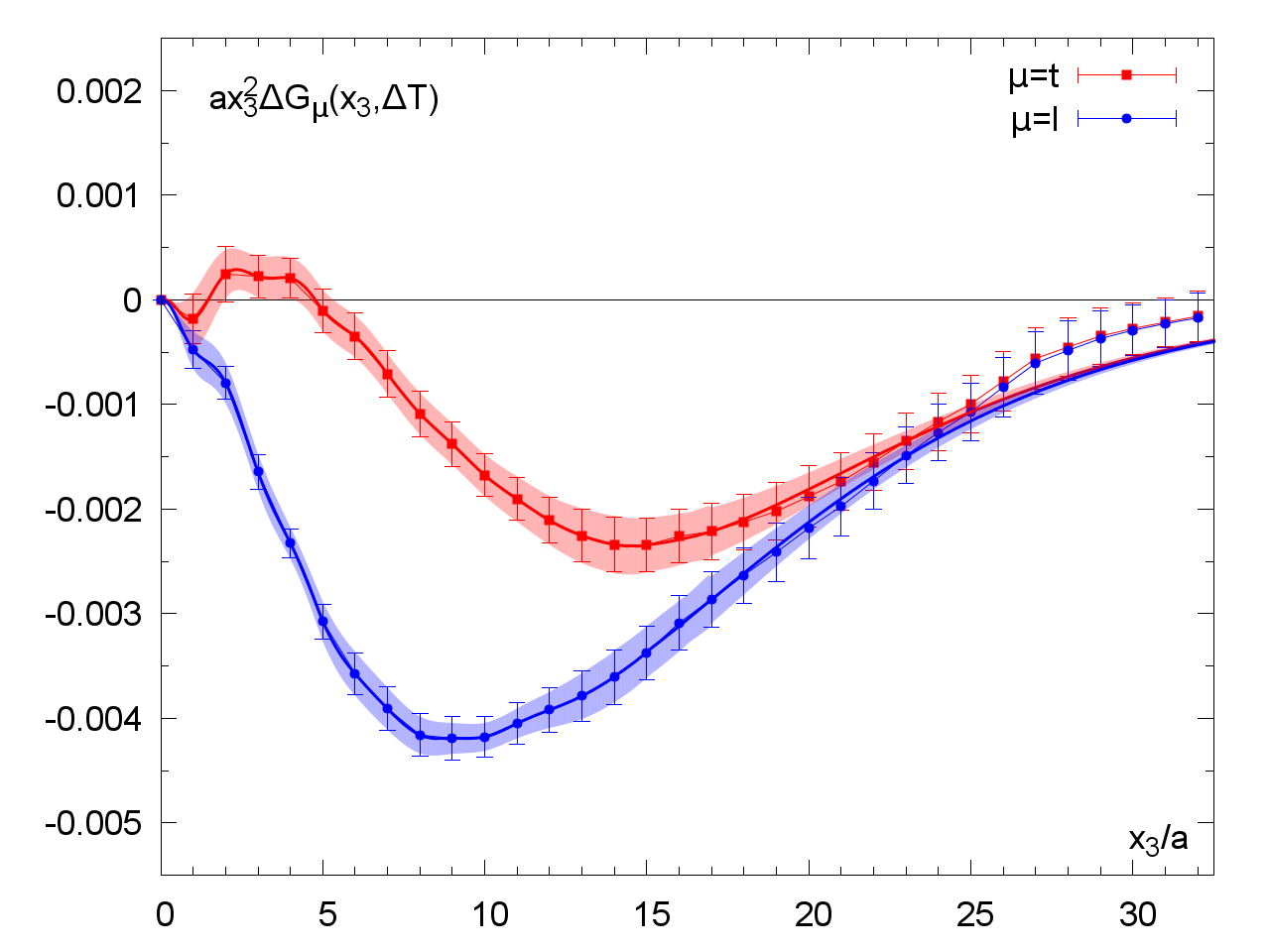}
\includegraphics[width=.5\textwidth]{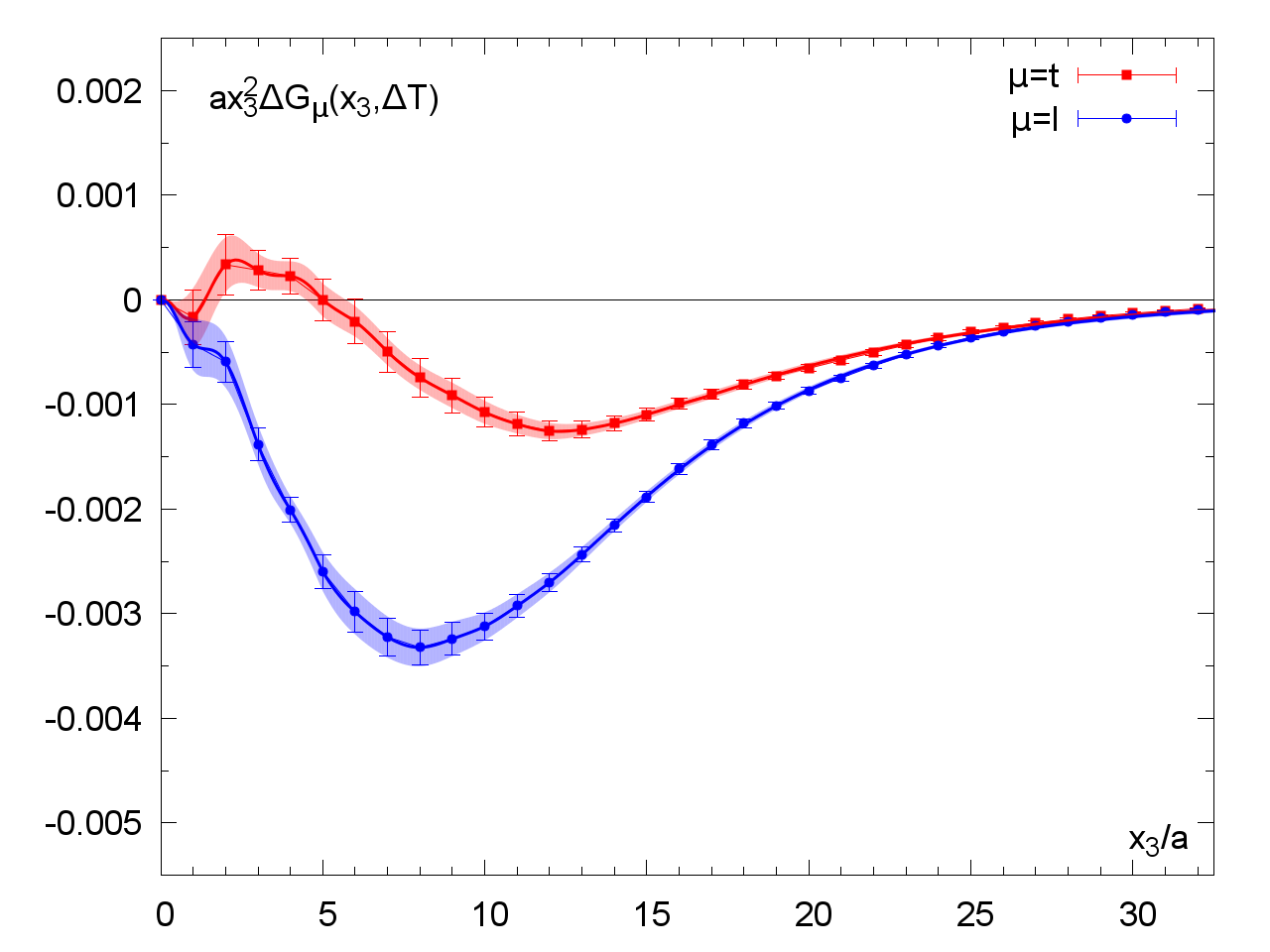}}
\caption{{The integrands $x_3^2\,\Delta G_\mu(x_3, T)$ 
for the convolution integrals of Eq.\ (\ref{eq:Gkappa1}) and Eq.\ (\ref{eq:Gkappa2}) 
for the light quark mass $m_{\rm ud}$ in the left panel and the strange quark mass $m_{\rm s}$ in the right panel.}}
\label{fig:integrand1}
\end{figure}

In order to obtain $\kappa_t$, $\kappa_l$ and $(\kappa_l-\kappa_t)$
given the local-conserved correlation functions of
Fig.\ \ref{fig:correlators}, one has to compute the following integrals:
\begin{align}
\la{eq:Gkappa1}
\kappa_t &= -\int_0^\infty dx\, x^2\,\Delta G_t(x,T),
 &\Delta G_t(x,T) \equiv G_{t}(x,T) - G(x,0), \\ 
\la{eq:Gkappa2}
\kappa_l &= -\int_0^\infty dx\, x^2\,\Delta G_l(x, T),
 &\Delta G_l(x, T) \equiv G_{l}(x,T) - G(x,0), \\  
\la{eq:Gkappa3}
\kappa_l - \kappa_t &= -\int_0^\infty dx\, x^2\,\Delta G_{l-t}(x,T),
 & \Delta G_{l-t}(x, T) \equiv G_{l}(x,T) - G_{t}(x,T).
\end{align}
To compute these integrals, we employ the parametrized
correlation functions of Fig.\ \ref{fig:correlators}, form the
relevant differences, multiply the latter by $x^2$ and 
integrate them. The  integrands for $\kappa_t$
and $\kappa_l$ in the light quark case are shown in
the left panel of Fig.\ \ref{fig:integrand1}, while the corresponding strange quark
results are given in the right panel.
At large distances, $x_3\gtrsim 2/T$, the integrand is strongly suppressed.
At the same time, the explicit factor $x_3^2$ 
induces an exact zero at $x_3=0$ in the integrand. For the longitudinal case a 
broad peak in the negative $y$-direction emerges with a minimum at
$x_3\simeq (2T)^{-1}$. In the transverse case the integrand first yields
positive results before passing through zero and also exhibiting a
negative peak shape. Both results depend strongly on the mass value, as
the light-quark curve drops roughly a factor 1.5--2.0 lower
than the strange-quark curve. 

The integral is carried out using standard numerical integration techniques
(for example the `global adaptive' strategy using the
`Gauss-Kronrod rule'  or `trapezoidal rule' options supplied by the
Mathematica-package work very well here) for a set of jackknife bins,
yielding a central value and an error estimate. We obtain, for the two quark masses,
\begin{align}
\kappa_t(m_{\rm ud})=0.0400(39)  &,\quad\quad\quad \kappa_t(m_{\rm s})=0.0169(17)  \,,\\
\kappa_l(m_{\rm ud})=0.0750(34)  &,\quad\quad\quad \kappa_l(m_{\rm s})=0.0458(19)  \,.
\end{align}
These values are similar to the corresponding one-loop QED predictions
for $m/T\simeq1$ (if one divides the former by the $N_c=3$ factor).
This reflects the fact that the infrared behavior of the vector
correlator in QCD is very different from its QED counterpart due to
confinement and chiral symmetry breaking.  In QED, a substantial
fermion mass mimicks to a certain extent the rapid fall-off of the QCD
correlator.

\begin{figure}[t]
\centering
\includegraphics[width=.6\textwidth]{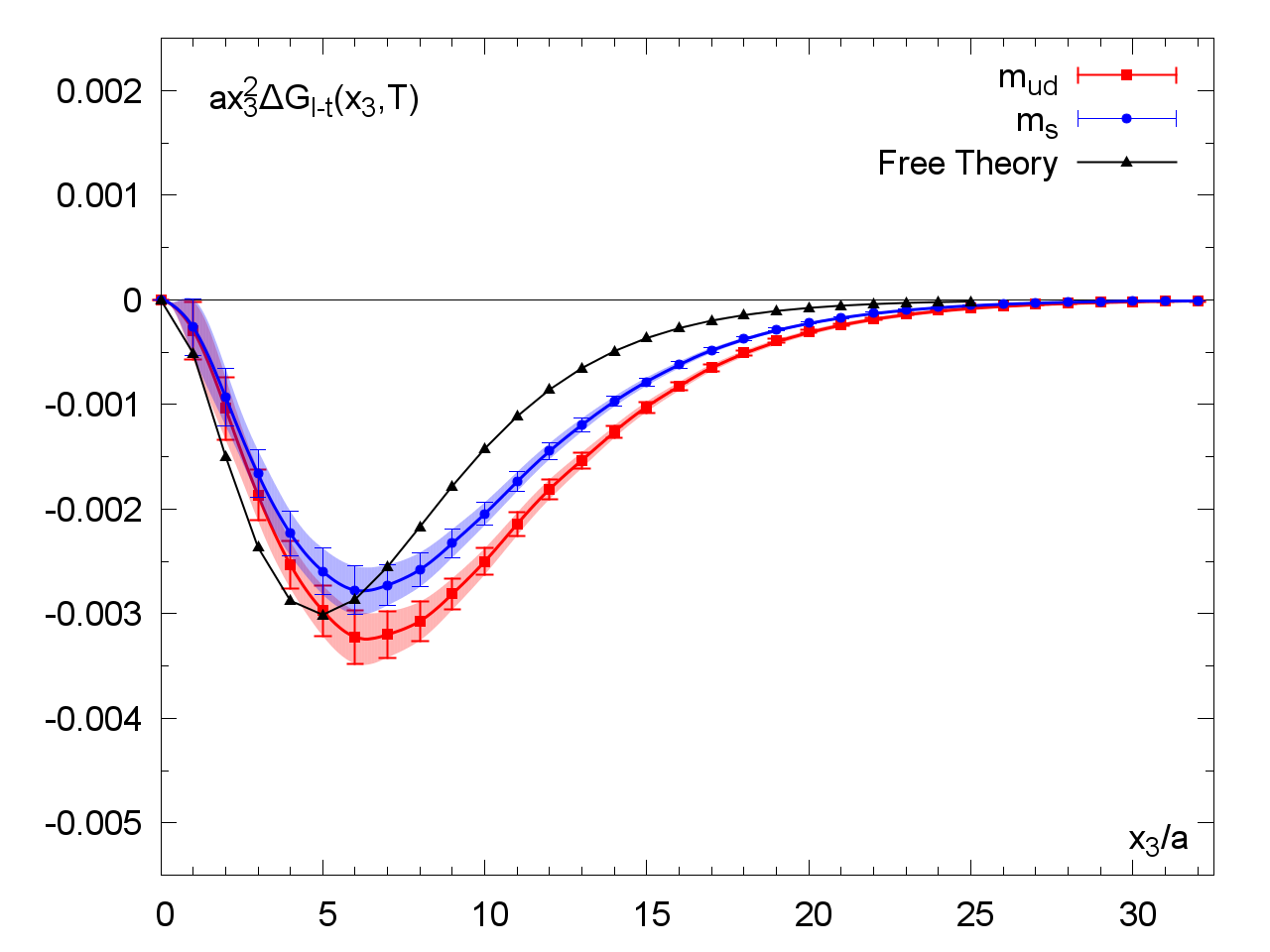}
\caption{{The integrand $x_3^2\,\Delta G_{l-t}(x_3,T)$ 
of Eq.\ (\ref{eq:Gkappa3}) for the light and the strange quark cases.
The corresponding quantity for free massless fermions is displayed as a black curve.}}
\label{fig:integrand2}
\end{figure}

Since the zero temperature correlator drops out in  $(\kappa_l - \kappa_t)$,
the latter difference is determined more accurately.
The corresponding integrand for both
masses is shown in Fig.\ \ref{fig:integrand2}, where we also give the
one-loop lattice result (computed in appendix \ref{sec:apdxlat}) for comparison. 
For both values of the quark mass we observe a broad peak in the negative direction. 
While qualitatively similar, the non-perturbative data 
falls off somewhat more slowly than the one-loop result, in spite of the fact that 
the quark mass is finite in the simulation and set to zero in the one-loop curve.
The resulting values of the integrals are:
\begin{align}
(\kappa_l - \kappa_t)(m_{\rm ud})=0.0350(24)  &,\quad\quad\quad (\kappa_l - \kappa_t)(m_{\rm s})=0.0290(21)\,.
\end{align}
Due to the slower fall-off of the screening correlators observed in the data, the value of 
this difference is noticibly larger than the one-loop calculation predicts, 
\begin{align}
\qquad\qquad (\kappa_l - \kappa_t)_{m/T=0.033}=0.0253  &,\quad\quad\quad (\kappa_l - \kappa_t)_{m/T=0.38}=0.0246\,
\qquad \quad (\textrm{one loop}),
\end{align}
particularly for the light quark mass case.
It would be interesting to see whether a two-loop calculation could account for the discrepancy.

\subsection{Implications for the free energy of two static leptons in the quark-gluon plasma}

To compute the free energy of two static electric charges in a QCD
plasma in the one-photon exchange approximation, it is useful to switch to a
position-space representation of the one-loop correction to the static
photon propagator. The reason is that the integral over the module of
the spatial momentum in Eq.\ (\ref{eq:F2gen}) is not absolutely
convergent.  Defining $\tilde j_\mu(x_3)\equiv \int dx_0 dx_1 dx_2
j_\mu(x)$, the representation
\ba \la{eq:freeen1}
F_l^{(2)}(r,T) &=& \frac{Q_1 Q_2 e^2\, e^{-m_{\rm el} r}}{4\pi r}\Big( 1+ e^2 h_l(r,T)\Big), \\
\la{eq:freeen2}
 h_l(r,T) &=&  \int_0^\infty dx_3\, \left(\phi_\ell(r,x_3)\,  \Big\< \tilde j_0(x_3)j_0(0)\Big\>_T 
 -  x_3^2\, \Big\< \tilde j_0(x_3)j_0(0)\Big\>_0\right), \\
\phi_\ell(r,x_3) &=& \left\{ \begin{array}{l@{\qquad}l} 
x_3^2 & x_3< r, \\
2x_3 r - r^2 & x_3\geq r ,
      \end{array}\right.
\ea
is equivalent, up to higher order terms in the coupling $e$, to
\eq(\ref{eq:F2gen}).  However, the integral over $x_3$ to be performed
numerically is now exponentially convergent for all $r$ in the
infrared, and the integrand is finite around $x_3=0$. Moreover, since
we identified the exponential fall-off of the correlator at long
distances, this representation is less affected by the finite box
length $L_3$ used in the simulation~\cite{Francis:2013fzp} than the
momentum space representation.

We recall that in the continuum, the two integrals contributing to
$h_{l}$ taken separately are logarithmically divergent in the
ultraviolet, but that their difference is finite.  On the lattice, the
corresponding sums are of course separately finite.  For large $r$,
$h_l(r,T)$ tends to $\kappa_l$.  For small $r$, the thermal correlator
can be approximated by the vacuum correlator and one recovers from
\eq(\ref{eq:freeen1}) the standard expression for the potential
$V(\vec r)$ modified by vacuum polarization effects, the so-called
Uehling potential (see for instance~\cite{Weinberg:1995mt}).

Similarly, in the case of two stationary, parallel currents, the
negative derivative of the free energy reads
\begin{figure}[t]
\centerline{\includegraphics[width=.5\textwidth]{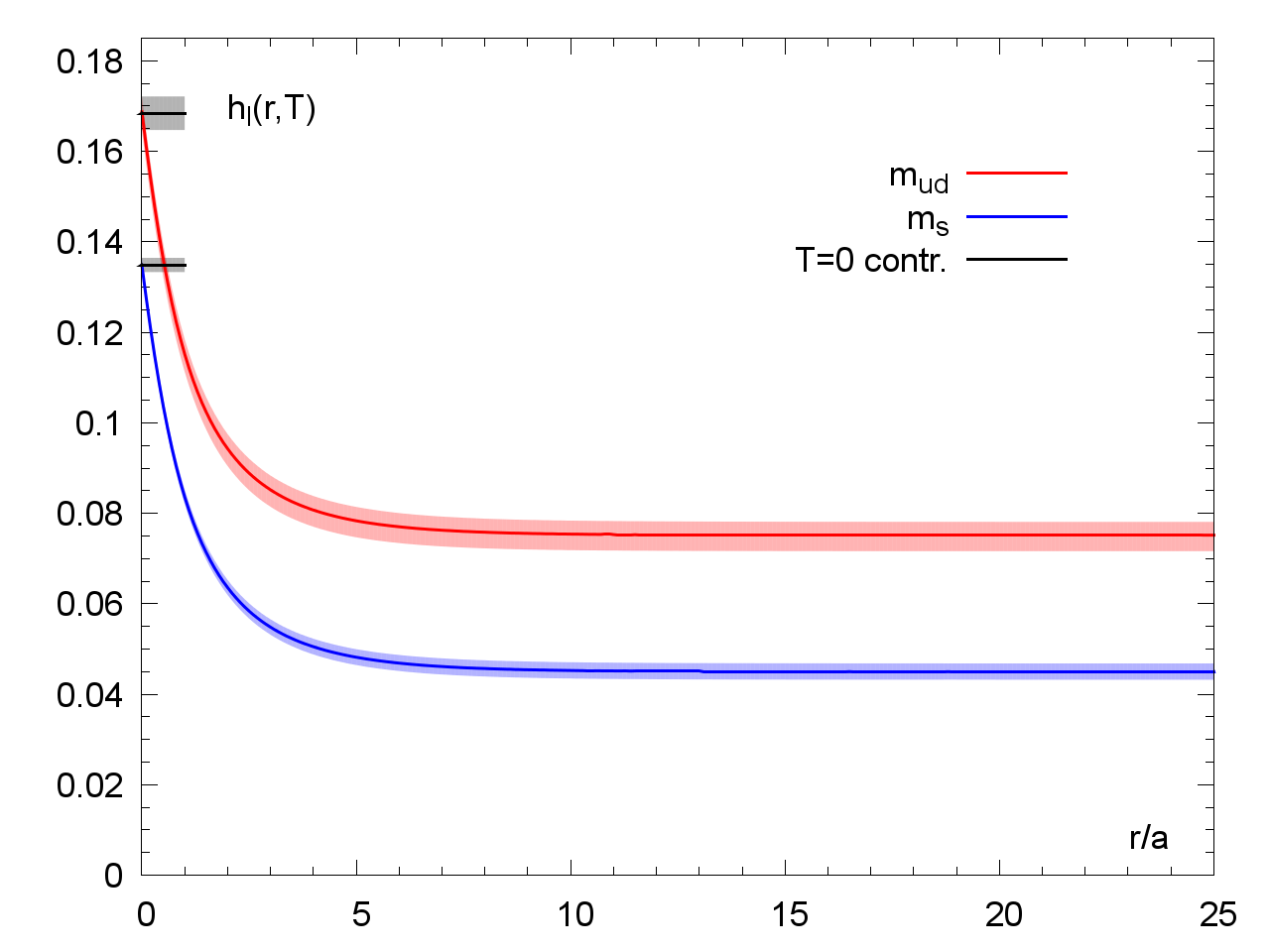}
\includegraphics[width=.5\textwidth]{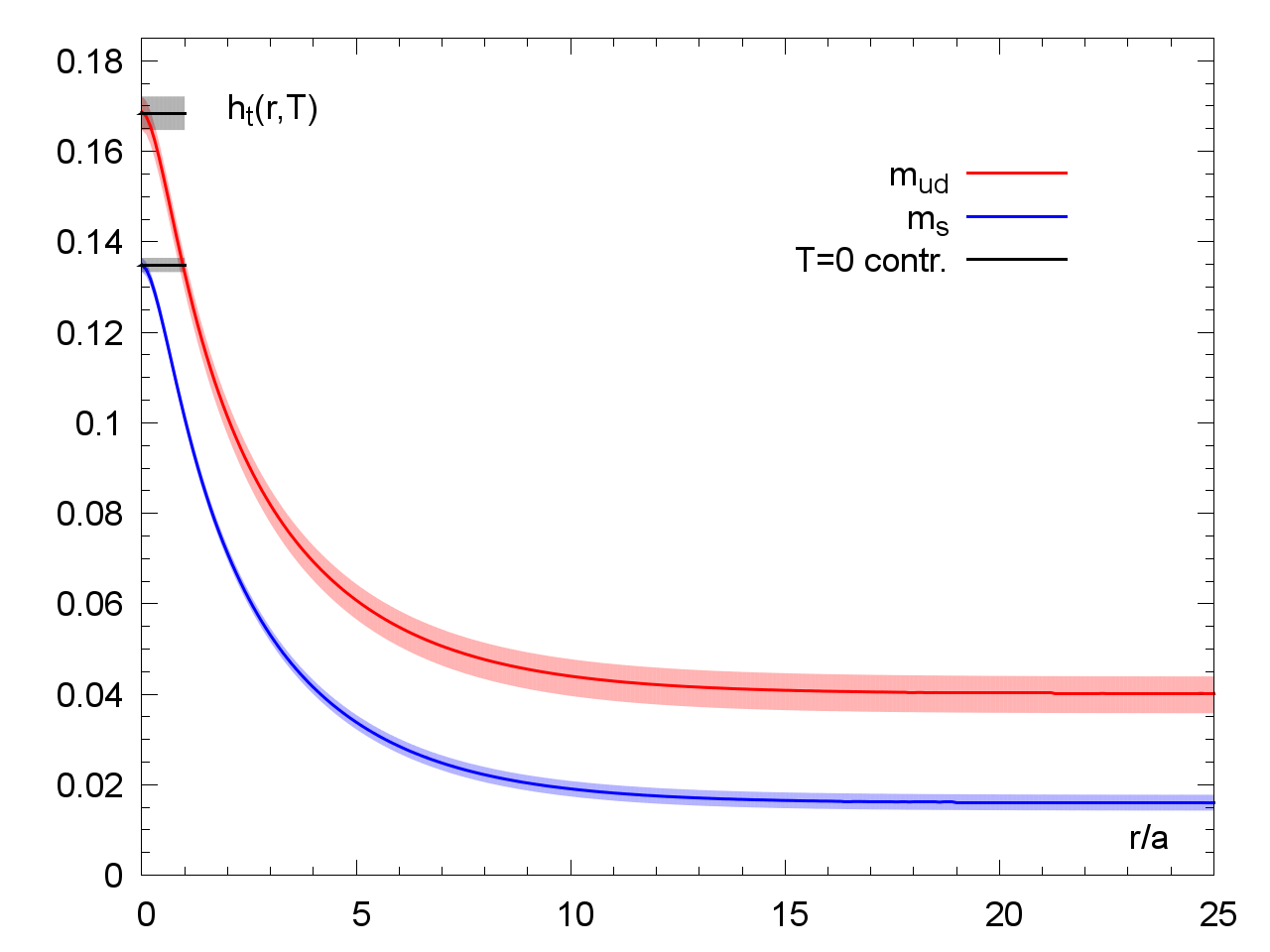}}
\caption{{The modification terms of the free energy of two static charges in a QED plasma. 
Left: $h_l(r,T)$ of Eq.\ (\ref{eq:freeen1}). Right: $h_t(r,T)$ of Eq.\ (\ref{eq:negdev1}). 
In both cases the light and strange quark results are displayed respectively in red and blue. 
We highlight the $r$-independent $T=0$ contribution in black around $r=0$.}}
\label{fig:potential}
\end{figure}
\ba \la{eq:negdev1}
 -\frac{1}{L_3}\frac{\partial}{\partial r} F_t^{(2)}(r,T) \! &=& \!
-\frac{I_1I_2 e^2}{2\pi r} \Big( 1 + e^2 h_t(r,T)\Big),\\
\la{eq:negdev2}
h_t(r,T) &=&
\int_0^\infty \! dx_3\, \left( \phi_t(r,x_3)\, \Big\<\tilde j_1(x_3)j_1(0)\Big\>_T
-  x_3^2\, \Big\<\tilde j_1(x_3)j_1(0)\Big\>_0 \right) ,
\\
\phi_t(r,x_3) &=& \left\{ \begin{array}{l@{\qquad}l}
x_3^2 
& x_3<r, \\
x_3^2 - x_3 \sqrt{x_3^2-r^2} - r^2 \log\frac{r}{x_3+\sqrt{x_3^2-r^2}}
& x_3\geq r.
    \end{array}\right.
\ea
These equations allow us to compute the generalization of the electric Amp\`ere force
for currents immersed in hot, strongly interacting matter.

Now the integrals in Eq.\ (\ref{eq:freeen1}) and
Eq.\ (\ref{eq:negdev1}) are of the same form as those of
Eqs.~(\ref{eq:Gkappa1}--\ref{eq:Gkappa3}) and to compute these loop
corrections we may reuse the machinery developed to calculate the
antiscreening coefficients. The results for both correction functions
are shown in Fig.\ \ref{fig:potential}, where the red and blue lines
denote the light and strange quark mass cases, respectively.  We give
the distance $r$ in lattice units and highlight the $r$-independent
$T=0$ contribution by short black bands around $r=0$ (again, this
contribution is logarithmically divergent in the continuum limit).  In
both cases we observe a sharp drop from the $T=0$ contribution at
small $r$ with the results leveling off to a constant.  In the case of
the free energy of two static leptons in the left panel of
Fig.\ \ref{fig:potential}, the dependence on $r$ rapidly becomes
negligible; it levels off into a constant shift to the leading order
term at around $r/a\simeq 10$. At this point the results for the two
masses differ by roughly a factor 1.5. In the case of the derivative
of the free energy of two parallel currents, the distance dependence
quickly becomes negligible and levels off around $r/a\simeq 20$. In
this case the mass dependence is seen to give roughly a factor 2
between the light and strange quark cases.  Comparing the errors from
the $T=0$ contribution and the respective $T\neq 0$ parts, we note
that the $T=0$ uncertainty dominates at large distances. In the limit
$r\rightarrow\infty$ the terms $h_l(r,T)$ and $h_t(r,T)$ become
equivalent to those of Eq.\ (\ref{eq:Gkappa1}) and
Eq.\ (\ref{eq:Gkappa2}), which provides a useful cross-check of our
calculations and indeed the results agree as expected.

\begin{figure}[t]
\centerline{\includegraphics[width=.5\textwidth]{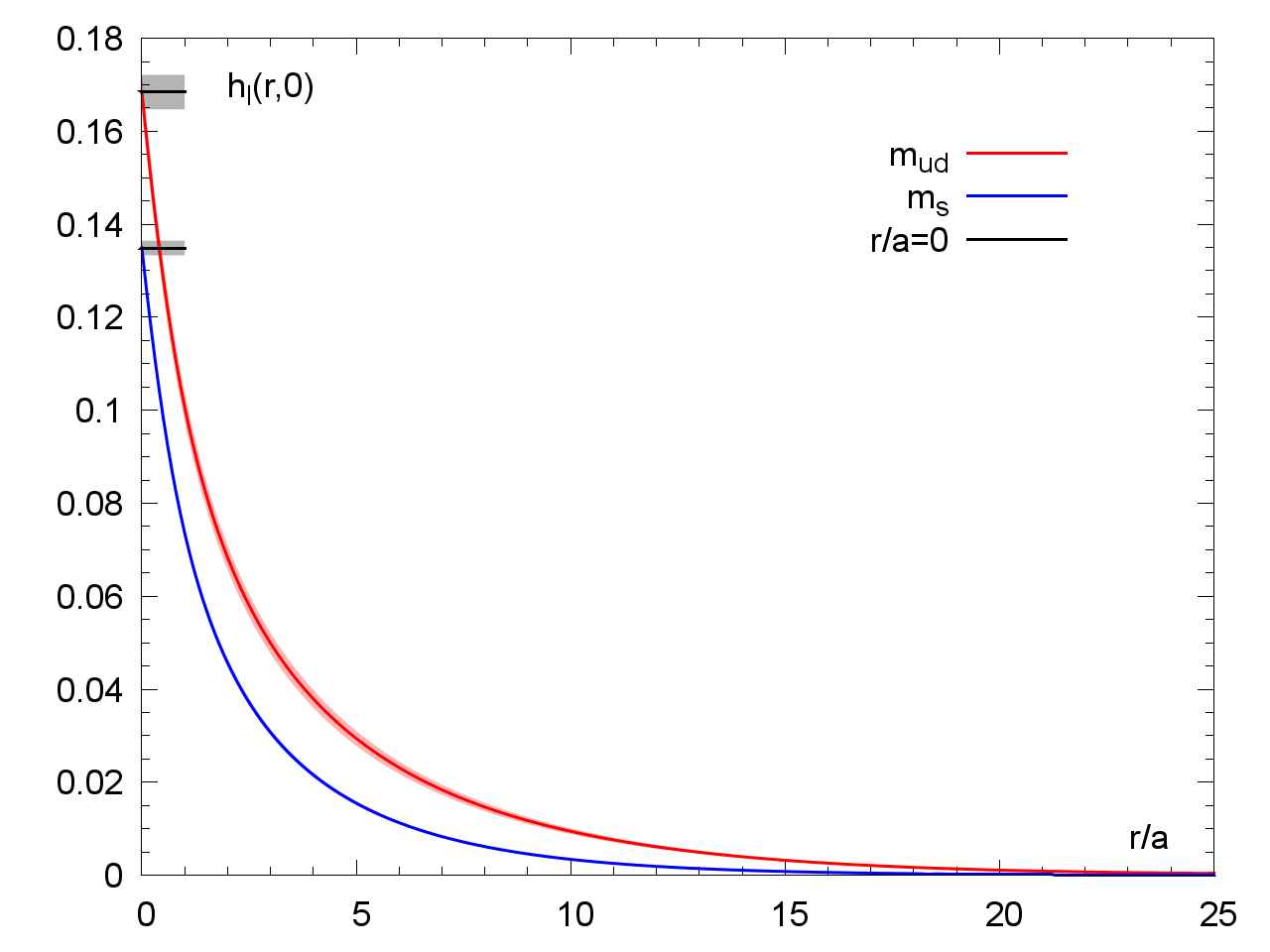}}
\caption{{The hadronic vacuum polarization contribution $h_l(r,0)$ to the potential between two electric charges. 
Note that $r/a=20$ corresponds roughly to $r=1$fm.}}
\label{fig:potential0}
\end{figure}

To illustrate the difference of this result with the zero-temperature case,
we display the hadronic vacuum polarization contribution to the electric potential between
two electric charges in Fig.\ \ref{fig:potential0}. In the limit $T\to0$, \eq (\ref{eq:freeen1}) simplifies
to
\ba
V(r) &=& \frac{Q_1 Q_2 e^2}{4\pi r}\Big( 1+ e^2 h_l(r,0)\Big), 
\\
h_l(r,0) &=& - \int_r^\infty dx_3\; (r-x_3)^2\; \<\tilde j_0(x_3)\,j_0(0)\>_0\,.
\ea
In particular, the relative correction $h(r,0)$ to the Coulomb potential is positive definite,
i.e.\ the effective QED coupling becomes stronger at short distances, as expected.
Unlike in the thermal case, the correction goes to zero rapidly beyond a distance of 1fm.

\section{Conclusion\la{sec:concl}}

We have studied two static properties of quantum relativistic plasmas.
They represent a loop correction to the free energy of external static
charges and stationary currents, where charge renormalization has been
performed.  In a hydrodynamic treatment, $\kappa_t$ enters the
constitutive equation of the current and thus governs the linear
response to an external magnetic field with a non-vanishing curl.  We
have calculated these two coefficients at one-loop order in perturbation
theory for QED and using lattice QCD for the non-Abelian
case. Remarkably, the presence of the plasma generates an
antiscreening of electric currents and thereby an enhancement of the
Amp\`ere force between two currents. Our emphasis has been on the
interpretation of these quantities and on the best suited strategies
to compute them; in order to make a realistic prediction for QCD it
would be necessary to explore the quark mass dependence and to take
the continuum limit of the coefficients $\kappa_{l,t}$.  While it
appears unlikely that these physical effects have any observable
consequence in primordial nucleosynthesis, it would be interesting to
investigate further whether the coefficient $\kappa_t$ has any
observable consequences, be it in the QED or in the QCD plasma.

\acknowledgments{ We are grateful for the access to the
  zero-temperature ensemble used here, made available to us through
  CLS.  We also warmly thank Georg von Hippel who provided the smeared vector
  correlator on this ensemble.  We acknowledge the use of computing
  time for the generation of the gauge configurations on the JUGENE
  computer of the Gauss Centre for Supercomputing located at
  Forschungszentrum J\"ulich, Germany; the finite-temperature ensemble
  was generated within the John von Neumann Institute for Computing
  (NIC) project HMZ21. The correlation functions were computed on the
  dedicated QCD platform ``Wilson'' at the Institute for Nuclear
  Physics, University of Mainz. This work was supported by the
  \emph{Center for Computational Sciences in Mainz} as part of the
  Rhineland-Palatinate Research Initiative and by the DFG grant ME
  3622/2-1 \emph{Static and dynamic properties of QCD at finite
    temperature}.}

 \appendix 

\section{Treelevel calculation of $\kappa_\ell$ and $\kappa_t$ \la{sec:apdxPT}}

With $\tilde j_1(x_3) = \int dx_0 dx_1 dx_2 j_1(x)$, the coefficient $\kappa_t$ can be computed as follows,
\be
\kappa_t = - \frac{\partial}{\partial(q_3^2)}\int_{-\infty}^\infty dx_3 \<\tilde j_1(x_3) j_1(0)\>|^T_0\; e^{iq_3x_3}\Big|_{q_3=0}
= +\frac{1}{2} \int_{-\infty}^\infty dx_3\; x_3^2  \<\tilde j_1(x) j_1(0)\>|^T_0\;.
\ee
Using the spectral representation of the screening correlator,
\be
- \<\tilde j_1(x_3) j_1(0)\> = \int_0^\infty \frac{d\omega}{2\pi} \; \tilde \rho_{11}(\omega,T) \; e^{-\omega |x_3|},
\ee
we obtain the expression
\be
\kappa_t = -\frac{1}{\pi}\int_0^\infty  \frac{d\omega}{\omega^3} \;(\tilde\rho_{11}(\omega,T)- \omega^2\rho^{\rm vac}(\omega^2)).
\ee
Introducing the notation
\be
p_n^{\pm} = (2n\pm 1)\pi T, \qquad E_n^\pm = (p_n^\pm{}^2 + m^2)^{1/2},
\ee
the spectral functions are given by, for $\omega\geq 0$,
\ba
\rho^{\rm vac}(\omega^2) &=& \frac{1}{6\pi} \sqrt{1-4m^2/\omega^2}\; (1+2m^2/\omega^2)\;  \theta(\omega-2m),
\\
\tilde \rho_{11}(\omega,T) &=& \frac{\omega}{2 \beta} 
\sum_{n=1}^\infty
 \left(1+\left({2E_n^-}/{\omega}\right)^2\right)\; \theta\left(\omega - 2 E_n^-\right).
\ea
We consider different intervals of frequency separately.
From the low-frequency region, we get a purely vacuum contribution,
\be\la{eq:kaplow}
\kappa^{\rm low} =  \frac{1}{\pi}\int_0^{2E_0^+} \frac{d\omega}{\omega} \rho^{\rm vac}(\omega^2)
= \frac{1}{36\pi^2}\left( 6\,{\rm asinh\,}(\pi/(\beta m)) - \frac{\pi (6\beta^2m^2 + 5\pi^2)}{(\beta^2 m^2 + \pi^2)^{3/2}}\right)
\ee
If we define $\tilde\rho_{11}(\omega,n,T)$ for $n\geq 1$ such that 
\be
\tilde\rho_{11}(\omega,n,T) = \tilde \rho_{11}(\omega,T) \qquad {\rm for}\qquad 
2E_n^- \leq \omega \leq 2E_n^+ ,
\ee
one finds 
\be
\tilde\rho_{11}(\omega,n,T) = \frac{n}{6\beta^3  \omega} 
\left(3\beta^2  (\omega^2 +4m^2) + 4\pi^2(4n^2-1) \right).
\ee
Thus the coefficient $\kappa_t$ is given by
\ba\la{eq:kapsep}
\kappa_t &=& \kappa^{\rm low} + \sum_{n=1}^\infty g_b(\beta,n),
\\
g_b(\beta,n) &\equiv& -\frac{1}{\pi}\int_{2E_n^- }^{2E_n^+ }
\frac{d\omega}{\omega^3} \; (\tilde \rho_{11}(\omega,n,T) -  \omega^2\rho^{\rm vac}(\omega^2))
\ea
The summand $g_b(\beta,n)$ is easily obtained analytically but 
the expression is not very illuminating and we do not reproduce it here.
The series is absolutely convergent. Moreover, $g_b(\beta,n)$ can be 
expanded in positive powers of $\beta$, and each term in the expansion
is an absolutely convergent series in $n$. The small $\beta m$ expansion
is thus straightforwardly obtained in this way.
Note that the infrared divergence in $\kappa_t$ when $m\to 0$ comes entirely from $\kappa^{\rm low}$.

For $\kappa_\ell$, the calculation proceeds in the same way, 
\be
\kappa_\ell = -\frac{1}{\pi}\int_0^\infty  \frac{d\omega}{\omega^3} (\tilde\rho_{00}(\omega,T) - \omega^2\rho^{\rm vac}(\omega^2)),
\ee
with the relevant screening spectral function given by
\be
\tilde\rho_{00}(\omega,T) = \frac{\omega}{\beta} \sum_{n=1}^\infty
\;(1-(2p_n^-/\omega)^2)\; \theta\left(\omega - 2 E_n^-\right).
\ee
We then have
\be
\tilde \rho_{00}(\omega,T) =\tilde \rho_{00}(\omega,n,T) \equiv \frac{ n \,\omega}{\beta } + \frac{ 4\pi^2 n (1-4n^2)}{3\beta^3 \omega},
\qquad \quad 2E_n^- <\omega <2E_n^+.
\ee
Defining $g_\ell(\beta,n)$ analogously to $g_b(\beta,n)$ with $\tilde\rho_{11}$ replaced
 by $\tilde \rho_{00}$, we have $\kappa_\ell = \kappa^{\rm low} + \sum_{n=1}^\infty g_\ell(\beta,n)$.

The expansion (\ref{eq:kp-kbexp}) for
$\kappa_\ell-\kappa_t=\sum_{n=1}^\infty [g_\ell(\beta,n) -  g_b(\beta,n)]$, is obtained from
\be
g_\ell(\beta,n) -  g_b(\beta,n) =  \frac{(p_n^+ + p_n^-) \, (E_n^+-E_n^-)}{48\pi^2 (E_n^- E_n^+)^3}
\left[(m^2+p_n^+p_n^-)\, ({E_n^+}^2 + E_n^+ E_n^- + {E_n^-}^2)\, - 3 {E_n^-}^2 {E_n^+}^2\right].
\ee
This expression can be expanded in positive powers of $\beta$, and then
the individual Taylor coefficients can be summed over $n$. The series
in $n$ are all absolutely convergent.

In order to obtain a representation more suitable for large values of $\beta m$,
one uses the Poisson summation formula to rewrite the screening correlators
\be
- \< \tilde j_\mu(x_3) j_\mu(0)\> = 2\sum_{j\in\mathbb{Z}} (-1)^j \int \frac{d^3\vec p}{(2\pi)^3}\; 
\frac{e^{-2E_{\vec p}|x_3|+ip_3\beta j}}{E_{\vec p}^2} (E_{\vec p}^2 -p_\mu^2),
\qquad \textrm{(no summation over $\mu$; $\mu\neq 3$)}.
\ee
In particular one obtains $\kappa_\ell-\kappa_t$ by integrating the difference of the two correlators 
over $x_3$,
\ba
\kappa_\ell - \kappa_t &=& \int_0^\infty dx_3\,x_3^2\; \left\<\tilde j_0(x_3) j_0(0) - \tilde j_1(x_3) j_1(0)\right\>
\\ &=& 
-\frac{1}{8\pi^2} \sum_{j\in\mathbb{Z}\backslash \{0\}} (-1)^j \int_0^\infty \frac{p\, dp}{E_p^5}\; \left(p^2 + 3\frac{\partial^2}{j^2\partial \beta^2}\right) 
\frac{\sin(p\beta j)}{\beta j}.
\ea
Note that the term $j=0$ does not contribute, because it corresponds
to the $T=0$ situation, where by rotation invariance the two
correlators are equal. The integrals are now linear combinations of modified Bessel functions and we arrive at \eq(\ref{eq:kp-kbexp}).

\section{Lattice perturbation theory\la{sec:apdxlat}}

We consider a fermion described by the Wilson action.
Defining 
\ba
\hatp_\mu = \frac{2}{a} \sin\frac{ap_\mu}{2},\qquad \quad 
\circp_\mu = \frac{1}{a} \sin ap_\mu, \qquad \quad M_p = \frac{1}{2}a\hatp^2 + m,
\\
\int_p f(p)\equiv \frac{1}{\beta} \sum_{n_0=1}^{\beta/a} 
 \int_{-\frac{\pi}{a}}^{\frac{\pi}{a}}\frac{d^3\vec p}{(2\pi)^3}
~f({(2n_0-1)\pi}/{\beta},\vec p),
\ea
the propagator reads
\be\la{eq:GwFourier}
\Gw(x,y) \equiv \<\psi(x) \bar\psi(y)\> = \int_p
\frac{e^{ip(x-y)}}{i\gamma_\mu \circp_\mu + M_p}.
\ee
Consider the local vector current $\bar\psi(x)\gamma_\mu\psi(x)$ and the conserved vector current
\be
V_\mu(x) = \frac{1}{2} \left( \bar\psi(x+a\hat\mu) (1+\gamma_\mu) U_\mu^\dagger(x)\psi(x)
 - \bar\psi(x)(1-\gamma_\mu) U_\mu(x) \psi(x+a\hat\mu)\right).
\ee
Its correlation function with the local current is
\ba
G^{\rm CL}_{\mu\nu}(x) &\equiv& 
 - \< V_\mu(x) \; \bar\psi(0)\gamma_\nu\psi(0)\>
\\
&=& \frac{1}{2} \tr\{(1+\gamma_\mu) G(x,0)\gamma_\nu G(0,x+a\hat\mu) 
 - (1-\gamma_\mu) G(x+a\hat\mu) \gamma_\nu G(0,x) \}.
\nonumber
\ea
Inserting expression (\ref{eq:GwFourier}) for the fermion propagator and performing the 
Dirac traces, we obtain
\ba\la{eq:PiCL}
&& a^4 \sum_x G^{\rm CL}_{\mu\nu}(x) \, e^{ik(x+a\hat\mu/2)} \equiv -\Pi^{\rm CL}_{\mu\nu}(k) 
\\ && 
= {4} \int_p
\frac{\cos[a(p_\mu+\frac{k_\mu}{2})] (\delta_{\mu\nu} (\circp\cdot\circq + M_p M_q) - \circp_\mu \circq_\nu - \circp_\nu \circq_\mu)
      - \sin[a(p_\mu+\frac{k_\mu}{2})] (M_p \circq_\nu + M_q \circp_\nu)}
{({\circp}^2 + M_p^2)\;({\circq}^2 + M_q^2)} \Big|_{q=p+k}.
\nonumber
\ea
The correlator satisfies 
\be
\sum_\mu \hatk_\mu \Pi^{\rm CL}_{\mu\nu}(k) = 0.
\ee

Another representation is, on a spatial torus with infinite time extent,
\ba
a^3 \sum_{\vec x}  G_{11}^{\rm CL}(x) &\stackrel{x_0\neq 0}{=}& 
\frac{4}{L_1L_2L_3} \sum_{\vec p}
\frac{e^{-2\omega_{\vec p} |x_0|}}{D(\vec p)^2}
\Big[\cos(a p_1) \left(D(\vec p)^2/(2A(\vec p))^2 + \vec{\circp}^2-2 \circp{}^2_1 + C(\vec p)^2\right)  -2a \circp{}^2_1 \,C(\vec p)\Big],\qquad 
\\
a^3 \sum_{\vec x}  G_{11}^{\rm CL}(0,\vec x) &=& \frac{4}{L_1L_2L_3} \sum_{\vec p}
\frac{1}{D(\vec p)^2}\Big[\cos(ap_1) \left(\vec{\circp}^2-2 \circp{}^2_1 + E(\vec p)^2 \right) -2a \circp{}^2_1 \, E(\vec p)\Big]
\ea
with 
\ba
A(\vec p) &=& 1 + am + \frac{1}{2} a^2 \hat{\vec p}^2,
\qquad 
B(\vec p) = m^2 + (1+am) \hat{\vec p}^2 + \frac{1}{2} a^2 \sum_{k<l} \hatp_k^2 \hatp_l^2.
\\
C(\vec p) &=& \frac{1}{2} a\hat{\vec p}^2 + m -\frac{aB(\vec p)}{2A(\vec p)},
\qquad 
D(\vec p) = \sqrt{B(\vec p)\,(4A(\vec p) + a^2 B(\vec p))},
\\
E(\vec p) &=& C(\vec p) + D(\vec p)/(2A(\vec p)),
\qquad
 \omega_{\vec p} = \frac{2}{a} {\rm \,asinh\,}\left( \frac{a}{2}\sqrt{B(\vec p)/A(\vec p)}\right).
\ea
This representation is useful to calculate the $T=0$ vector correlator
(in which case the $L_i$ can be sent to infinity) and the screening
correlator at finite $T$ (in which case the direction $\vec e_0$ is
interpreted as a spatial direction, while one of the directions $\vec
e_k$ is interpreted as the Matsubara cycle). For the correlator of two
local vector currents, see~\cite{Aarts:2005hg}.

\bibliography{/Users/harvey/BIBLIO/viscobib.bib}

\end{document}